\documentclass[aip,jcp,reprint,onecolumn,frontmatterverbose,groupedaddress]{revtex4-1}
\usepackage{graphicx}
\usepackage{dcolumn}
\usepackage{bm}
\usepackage[mathlines]{lineno}

\newcommand{\rank}{\text{rank}}

\begin{document}

\title[New exact solutions of 2DKK and 2DSK equations]{New exact solutions of two-dimensional integrable generalizations of Kaup-Kuperschmidt and Sawada-Kotera equations via the $\overline\partial$-dressing method.}

\author{V.G. Dubrovsky}
\email[E-mail:]{dubrovsky@ngs.ru}
\author{A.V. Topovsky}
\author{M.Yu. Basalaev}
\affiliation{Novosibirsk State Technical University, Karl Marx prosp. 20, Novosibirsk 630092, Russia.}

\date{\today}

\begin{abstract}
The 2+1-dimensional integrable generalization of Kaup-Kuperschmidt and Sawada-Kotera equations are studied by $\overline\partial$-dressing method of Zakharov and Manakov. The solutions  with functional parameters and periodic solutions are constructed.
\end{abstract}


\maketitle

\section{Introduction}

In the last three decades the Inverse Spectral Transform (IST) method has been generalized and successfully applied to various
2+1-dimensional nonlinear evolution equations such as Kadomtsev-Petviashvili, Davey-Stewardson, Veselov-Novikov,
Zakharov-Manakov system, Ishimory, two dimensional integrable sine-Gordon and others, see the books \cite{NZMP,AK,K1,K2} and references there. The nonlocal Rieman-Hilbert \cite{M}, $\overline\partial$-problem \cite{BC1} and more general $\overline\partial$-dressing method of Zakharov and Manakov \cite{ZM,Z1,BM} are now basic tools for solving 2+1-dimensional integrable nonlinear equations, see also the books \cite{NZMP,AK,K1,K2} and reviews \cite{FA,BC2,Z2}.

In the present paper the $\overline\partial$-dressing method is applied for the construction
of new classes of solutions with functional parameters and as their particular
cases periodic solutions
of 2+1-dimensional integrable generalizations of Kaup-Kuperschmidt  (2DKK),
\begin{equation}\label{2DKK}
u_{t} + u_{xxxxx} + \frac{25}{2} u_{x} u_{xx} + 5 u u_{xxx} + 5u^{2}u_{x} + 5u_{xxy} - 5 \partial_{x}^{-1} u_{yy} + 5 u u_{y} +
5u_{x}\partial_{x}^{-1} u_{y}=0;
\end{equation}
and Sawada-Kotera (2DSK),
\begin{equation}\label{2DSK}
u_{t} + u_{xxxxx} + 5 u_{x} u_{xx} + 5 u u_{xxx} + 5u^{2}u_{x} + 5u_{xxy} - 5 \partial_{x}^{-1} u_{yy} + 5 u u_{y} +
5u_{x}\partial_{x}^{-1} u_{y}=0,
\end{equation}
equations. These equations have been discovered in paper \cite{KD1}. Now it is well known that the Sawada-Kotera equation belongs to the BKP hierarchy, and the Kaup-Kupershmidt equation to the CKP hierarchy   \cite{DJK}. These equations can be represented as the compatibility conditions in the Lax form $[L_1,L_2] = 0$; for the 2DKK equation of the following two linear auxiliary problems \cite{KD1},
\begin{eqnarray}\label{auxiliary problems for 2DKK}
L_1 \psi = (\partial_{x}^3 +u \partial_{x} + \frac{1}{2}u_{x}+\partial_{y})\psi = 0, \\
L_2 \psi =[\partial_t - 9 \partial_x^5 - 15u\partial_x^3 - \frac{45}{2} u_x \partial_x^2-
(\frac{35}{2}u_{xx}+5u^2- 5 \partial_x^{-1}u_y) \partial_x-(5uu_x-\frac{5}{2}u_y + 5u_{xxx})]\psi = 0;
\end{eqnarray}
and for 2DSK equation of another two linear auxiliary problems \cite{KD1},
\begin{eqnarray} \label{auxiliary problems for 2DSK}
L_{1} \psi = (\partial_{x}^3 +u\partial_{x} +\partial_{y})\psi = 0,  \\
L_{2} \psi = [\partial_t - 9 \partial_x^5 - 15 u\partial_x^3-15u_x \partial_x^2- (10u_{xx}+5u^2- 5 \partial_x^{-1}u_y)
\partial_x]\psi = 0.
\end{eqnarray}
Here and bellow $\partial_x \equiv \partial/\partial_x,\ldots$ and $\partial_x^{-1}$ is an operator inverse to $\partial_x$.

The first linear auxiliary differential problems in (\ref{auxiliary problems for 2DKK}) and (\ref{auxiliary problems for 2DSK}) are of the third order on $\partial_x$, such problems in
general position have several fields as the coefficients at the various degrees of $\partial_x$.
The 2DKK  (\ref{2DKK}) and 2DSK (\ref{2DSK}) equations  arise as special reductions of some integrable nonlinear systems for these fields. It is well known that the study of special reductions requires more attention and may be more difficult than the consideration of nonlinear equations integrable by auxiliary linear problems in general
position.

The scheme for construction of solutions with functional parameters for 2+1-dimensional integrable equations by the example of KP equation was developed earlier in the famous papers \cite{ZakhShab_74,ZakhShab_79}  of Zakharov and Shabat in the framework of their's variant of dressing method, see also in this connection the book \cite{NZMP}.

The present paper is natural continuation of the paper \cite{DubrLisit} of the first author. In the paper \cite{DubrLisit} $\overline\partial$-dressing method of Zakharov and Manakov was at first used  for construction of multiline soliton solutions of 2DKK and 2DSK equations; some line solitons of considered equations were constructed earlier by another means, see for example the paper \cite{Xing}. The application of $\overline\partial$-dressing method in nonstandard situations, in our case some nonlinear constraints on the wave functions of the linear auxiliary problems must be satisfied as special reductions, may be very useful and instructive. Let us underline that all of our constructions of exact solutions of considered equations are based exclusively on $\overline\partial$-dressing method and not depend on the relations of 2DKK, 2DSK equations with CKP and BKP hierarchies correspondingly.

Our paper is organized as follows. In the section II the basic ingredients of $\overline\partial$-dressing method are shortly reviewed and some useful formulas derived in the paper \cite{DubrLisit} for 2DKK and 2DSK equations (\ref{2DKK}) and (\ref{2DSK}) are presented: reconstruction formulas,  nonlinear constraints on the wave functions,  the conditions of reality of  solutions and so on.
The new classes of exact solutions with functional parameters  for the 2DKK and 2DSK equations   are constructed correspondingly in sections III and IV. In section V as a particular cases of solutions with functional parameters the periodic solutions of 2DKK and 2DSK equations are calculated. Section VI contains some conclusions and acknowledgements.

\section{Basic formulas of  $\overline\partial$-dressing method for  2DKK and 2DSK equations.}

In this section for  convenience we are going to remind some useful general formulas of  $\overline\partial$-dressing method for  2DKK and 2DSK equations (\ref{2DKK}) and (\ref{2DSK}), see the paper \cite{DubrLisit} for more details.

At first one postulates  non-local $\overline\partial$-problem \cite{ZM,Z1,BM},
\begin{equation}\label{dibar problem}
\frac{\partial\chi}{\partial\overline\lambda} = (\chi* R)(\lambda,\overline\lambda) = \int\int_{C} d\mu
\wedge d\overline\mu \chi(\mu;\overline\mu)R(\mu,\overline{\mu};\lambda,\overline\lambda),
\end{equation}
here $\chi$ and $R$ in considered case are scalar complex valued functions. For  wave function $\chi$ we choose  canonical normalization: $\chi\rightarrow 1$ as $\lambda\rightarrow\infty$. We assume also that  problem (\ref{dibar problem}) is uniquely solvable.
The solution of  $\overline{\partial}$-problem (\ref{dibar problem}) with constant normalization is equivalent to  solution of the following singular integral equation:
\begin{equation}\label{di_problem1}
\chi (\lambda) = 1 + \int\int\limits_C {\frac{d{\lambda }'\wedge
d{\overline {\lambda'}}}{2\pi i(\lambda'-\lambda)}}
\int\int\limits_C  \chi(\mu,\overline{\mu})
R(\mu ,\overline \mu ;\lambda',\overline {\lambda' }){d\mu \wedge d\overline{\mu }}.
\end{equation}

Then one introduces the dependence of a kernel $R$ of $\overline\partial$-problem (\ref{dibar problem}) on space and time variables $x,y,t$ \cite{DubrLisit},
\begin{equation}\label{R&F}
R(\mu,\overline\mu;\lambda,\overline\lambda;x,y,t) =R_{0}(\mu,\overline\mu;\lambda,\overline\lambda)e^{F(\mu;x,y,t)-F(\lambda;x,y,t)}; \quad F(\lambda;x,y,t) := i(\lambda x + \lambda^3 y +9 \lambda^5 t).
\end{equation}
At the next stage of  $\bar{\partial}$-dressing method \cite{ZM,Z1} one construct auxiliary linear problems for  2DSK and 2DKK equations, which in general form are given by expressions \cite{DubrLisit},
\begin{eqnarray}\label{Aux operators in long derivatives 2DSK&2DKK}
  L_{1}\psi &=& (\partial_{y}+\partial_{x}^3+u\partial_{x}+v)\psi=0,\\
  L_{2}\psi &=& (\partial_{t} - 9\partial_{x}^5 + w_3 \partial_{x}^3 + w_2 \partial_{x}^2 +w_1\partial_{x} +w_0)\psi=0.\nonumber
\end{eqnarray}
The wave function $\psi$ in (\ref{Aux operators in long derivatives 2DSK&2DKK}) is connected with wave function $\chi$ by the relation $\psi: = \chi e^{F(\lambda;x,y,t)}$.

Reconstruction formulas for the potentials of problems (\ref{Aux operators in long derivatives 2DSK&2DKK}) are express these potentials through some coefficients of series expansions of wave function $\chi$ in terms of powers of spectral variable $\lambda$ near the points $\lambda=0$ and $\lambda=\infty$,
\begin{equation}\label{series of chi}
\chi =\chi_{0}+\chi_{1}\lambda+\chi_{2}\lambda^{2}+\ldots, \quad
\chi =\chi_{\infty}+\frac{\chi_{-1}}{\lambda}+
\frac{\chi_{-2}}{\lambda^{2}}+\ldots;
\end{equation}
these formulas for the potentials of the first linear problem (\ref{Aux operators in long derivatives 2DSK&2DKK}) have the forms \cite{DubrLisit},
\begin{equation}\label{V reconstructFormulae 2DSK&2DKK}
    v =-3i\chi_{-1xx} +3\chi_{-2x} - 3\chi_{-1}\chi_{-1x};\quad
    u=-3i\chi_{-1x}.
\end{equation}
The coefficients  $\chi_{-1}$ and $\chi_{-2}$ due to (\ref{di_problem1}) are given by expressions,
\begin{equation}\label{di_problem_chi_-1}
\chi_{-1}  = - \int\int \limits_C {\frac{d{\lambda }\wedge
d{\overline {\lambda }}}{2\pi i}}
\int\int\limits_C  \chi(\mu,\overline{\mu})
R_0(\mu ,\overline \mu ;\lambda
,\overline {\lambda })e^{F(\mu) -
F(\lambda)}{d\mu \wedge d\overline{\mu }},
\end{equation}
\begin{equation}\label{di_problem_chi_-2}
\chi_{-2}  = - \int\int \limits_C {\frac{\lambda d{\lambda }\wedge
d{\overline {\lambda }}}{2\pi i}}
\int\int\limits_C  \chi(\mu,\overline{\mu})
R_0(\mu ,\overline \mu ;\lambda
,\overline {\lambda })e^{F(\mu) -
F(\lambda)}{d\mu \wedge d\overline{\mu }},
\end{equation}
here and below, for abbreviation,  short notations $F(\lambda)$ for $F(\lambda;x,y,t)$ will be used.

It was shown in the paper \cite{KD1} that to  2+1-dimensional integrable generalizations of nonlinear Kaup-Kuperschmidt (\ref{2DKK}) and Sawada-Kotera (\ref{2DSK}) equations correspond  the reductions,
\begin{equation}\label{V for 2DKK,2DSK}
(2DKK):\qquad
v=\frac{1}{2}u_x;\qquad (2DSK):\qquad v=0.
\end{equation}
In terms of the wave function $\chi$ the reductions (\ref{V for 2DKK,2DSK}) can be expressed as following nonlinear constraints on the coefficients $\chi_{-1}$ and $\chi_{-2}$ \cite{DubrLisit},
\begin{equation}\label{reduction 2DKK}
(2DKK):\qquad \chi_{-2x} -\frac{i}{2}\chi_{-1xx} -\chi_{-1}\chi_{-1x}=0;
\end{equation}
\begin{equation}\label{reduction 2DSK}
(2DSK):\qquad \chi_{-2x} -i\chi_{-1xx} -\chi_{-1}\chi_{-1x}=0.
\end{equation}

Reconstruction formulas for the potentials of the second auxiliary problem (\ref{Aux operators in long derivatives 2DSK&2DKK}) due to
(\ref{V reconstructFormulae 2DSK&2DKK}) and reductions (\ref{V for 2DKK,2DSK}) have the forms \cite{DubrLisit},
\begin{eqnarray} \label{w for 2DKK}
(2DKK):\qquad
w_{1} = -\frac{35}{2}u_{xx}+5 \partial_x^{-1}u_{y} - 5u^2,\nonumber \\
w_{2} = -\frac{45}{2}u_{x},\quad w_{3}=- 15 u,\quad
w_{0} = \frac{5}{2}u_{y}-5 u_{xxx} - 5uu_{x},
\end{eqnarray}
in the case of 2DKK equation  (\ref{2DKK}) and,
\begin{eqnarray}\label{w for 2DSK}
(2DSK):\qquad
 w_{1} = -10 u_{xx}+5 \partial_{x}^{-1}u_{y} - 5u^2,\nonumber\\
 w_{2} = -15u_{x},\quad w_{3}=- 15 u,\quad
 w_{0} =0,
\end{eqnarray}
in the case of 2DSK equation  (\ref{2DSK}).

One can easily obtain from (\ref{V reconstructFormulae 2DSK&2DKK}) and (\ref{di_problem_chi_-1}) the restrictions on the kernel $R_0$ of  $\overline\partial$-problem (\ref{dibar problem}) following from reality of $u$; one derives in the limit of weak fields \cite{DubrLisit},
\begin{equation}\label{RC1}
R_0(\mu,\overline\mu;\lambda,\overline\lambda) =\overline{R_0(-\overline\mu,-\mu;-\overline\lambda,-\lambda)};\quad R_0(\mu,\overline\mu;\lambda,\overline\lambda)=\overline{R_0(\overline\lambda,\lambda;\overline\mu,\mu)}.
\end{equation}

The conditions of reductions (\ref{V for 2DKK,2DSK}) or (\ref{reduction 2DKK}), (\ref{reduction 2DSK}) and reality (\ref{RC1}) for $u$ lead to some restrictions on the kernel $R_0$ of $\overline{\partial}$-problem (\ref{dibar problem}) in the cases of the 2DSK and 2DKK equations.
It is evident that  conditions (\ref{RC1})  are the same for both 2DKK and 2DSK equations (\ref{2DKK}) and (\ref{2DSK}) but  nonlinear constraints (\ref{reduction 2DKK}) and (\ref{reduction 2DSK}) for these equations have different forms.
So in order to calculate  exact solutions of 2DKK (\ref{2DKK}) and 2DSK (\ref{2DSK}) equations via  $\overline\partial$-dressing method one must satisfy to conditions of nonlinear constraints (\ref{reduction 2DKK}),  (\ref{reduction 2DSK}) and reality (\ref{RC1}); this is main and difficult part of all constructions.

For degenerate kernel $R_0(\mu,\overline{\mu};\lambda,\overline{\lambda})$ of $\overline{\partial}$-problem (\ref{dibar problem}),
\begin{equation}\label{Kernel FP}
R_{0}(\mu,\overline{\mu},\lambda,\overline{\lambda})=
\pi\sum\limits_{k=1}^{N}f_k(\mu,\overline{\mu})g_k(\lambda,\overline{\lambda}),
\end{equation}
one can easily derive general determinant formula for the class of exact solutions $u(x,y,t)$ with functional parameters of  2DKK and 2DSK equations (\ref{2DKK}), (\ref{2DSK}).
Indeed, from (\ref{di_problem_chi_-1}), (\ref{di_problem_chi_-2}) and (\ref{Kernel FP}) follow compact formulas
for the coefficients $\chi_{-1},\chi_{-2}$ of the expansion (\ref{series of chi}) of function $\chi$,
\begin{equation}\label{chi_-1,-2_FP}
\chi_{-1}=-\frac{1}{2i}\sum\limits_{l,k=1}^{N}A^{-1}_{kl}\alpha_l\beta_k,\quad
\chi_{-2}= -\frac{1}{2}\sum\limits_{l,k=1}^{N}A^{-1}_{kl}\alpha_l\beta_{k\,x},
\end{equation}
where  the matrix $A$ has the form:
\begin{equation}\label{matrix1_FP}
A_{lk}=\delta_{lk}+\frac{1}{2}\partial_{x}^{-1}\alpha_l \beta_k.
\end{equation}

The functions $\alpha_k(x,y,t)$, $\beta_k(x,y,t)$ in (\ref{chi_-1,-2_FP}) and (\ref{matrix1_FP}),  which given by formulas,
\begin{equation}\label{alpha_FP,beta_FP}
\alpha_l(x,y,t): =\int\int\limits_C f_l(\mu,\overline{\mu})e^{F(\mu)}d\mu\wedge d\overline{\mu},\quad
\beta_l(x,y,t): = \int \int\limits_C g_l(\lambda,\overline{\lambda})
e^{-F(\lambda)}d\lambda \wedge d\overline{\lambda},
\end{equation}
are known as functional parameters (in coordinate representation). The functions $f_k(\mu,\overline{\mu}) , g_k(\lambda,\overline{\lambda})$ can be named as functional parameters in spectral representation.
 By  definitions (\ref{R&F}) and (\ref{alpha_FP,beta_FP}) the functional parameters $\alpha_n$ and $\beta_n$ satisfy to  following linear equations,
\begin{equation}\label{linearDiffEqForParam_FP}
\alpha_{n\,y}+\alpha_{n\,xxx}=0,\quad
\alpha_{nt}+\alpha_{n\,xxxxx}+5\alpha_{n\,xxy}-5\partial^{-1}_{x}\alpha_{n\,yy}=0;
\end{equation}
\begin{equation}\label{linearDiffEqForParam_FP}
\beta_{n\,y}+\beta_{n\,xxx}=0, \quad
\beta_{nt}+\beta_{n\,xxxxx}+5\beta_{n\,xxy}-5\partial^{-1}_{x}\beta_{n\,yy}=0.
\end{equation}

Here and below useful determinant identities,
\begin{equation}\label{useful_identities_FP}
Tr(\frac{\partial A}{\partial x}A^{-1})=\frac{\partial}{\partial x} \ln (\det A),\quad
Tr(B\;A^{-1})=\frac{\det(A+B)}{\det A}-1,
\quad 1+TrB=\det(1+B),
\end{equation}
will be used. The matrices  $B$ and $B\;A^{-1}$ in (\ref{useful_identities_FP}) are degenerate with rank 1.
With help of the first identity in (\ref{useful_identities_FP}) expression for $\chi_{-1}$  (\ref{chi_-1,-2_FP}) takes the form,
\begin{equation}\label{chi_-1_FP1}
\chi_{-1}=i\sum\limits_{k,l=1}^{N}A^{-1}_{kl}\frac{\partial A_{lk}}{\partial x}=iTr(A^{-1}\frac{\partial A}{\partial x})=
i\,\partial_{x}(\ln \det A).
\end{equation}
From (\ref{chi_-1_FP1}) by use of reconstruction formula (\ref{V reconstructFormulae 2DSK&2DKK}) one obtains general determinant formula for the solution $u$ with functional parameters $\alpha_k(x,y,t)$, $\beta_k(x,y,t)$ of  2DKK (\ref{2DKK}) and 2DSK (\ref{2DSK}) equations,
\begin{equation}\label{Solution_general_formula_FP}
u(x,y,t)=-3i\chi_{-1\,x}=3\frac{\partial^2}{\partial x^2}\ln \det A,
\end{equation}
here the elements of matrix $A$ are given by (\ref{matrix1_FP}).

In following sections III and IV  calculations of exact solutions $u(x,y,t)$ via simultaneous satisfaction to conditions of nonlinear reductions (\ref{reduction 2DKK}), (\ref{reduction 2DSK}) and  reality (\ref{RC1})   are performed for convenience  in cases of 2DKK (\ref{2DKK}) and 2DSK (\ref{2DSK}) equations separately; analogous problem of calculations of multiline soliton solutions for these equations was solved in the previous paper \cite{DubrLisit}.

\section{Solutions of  2DKK equation.}

In this section we calculate new classes of  solutions with functional parameters and as their particular cases reproduce multiline solitons \cite{DubrLisit}.

The reduction condition (\ref{reduction 2DKK}) imposes some restrictions on functional parameters $\alpha_l(x,y,t)$, $\beta_l(x,y,t)$ (\ref{alpha_FP,beta_FP}). Substituting coefficients $\chi_{-1}$ and $\chi_{-2}$ from (\ref{chi_-1,-2_FP}) into (\ref{reduction 2DKK}) the reduction condition for  2DKK equation can be rewritten in the form,
\begin{equation}\label{reduction1 2DKK}
\frac{\partial}{\partial x}\Big[\chi_{-2} -\frac{i}{2}\chi_{-1x} -\frac{1}{2}\chi_{-1}^2\Big]=
\sum^{N}_{k,l=1}\frac{\partial}{\partial x}\Big[\alpha_l\beta_{k\,x}-\alpha_{l\,x}\beta_{k}\Big]A^{-1}_{kl}=\frac{\partial}{\partial x}Tr (B A^{-1})=0,
\end{equation}
where matrix $B$ has elements,
\begin{equation}\label{matrix B}
B_{lk}=\alpha_l\beta_{k\,x}-\alpha_{l\,x}\beta_{k}.
\end{equation}
It will be shown below that reduction condition (\ref{reduction 2DKK}), or
equivalently (\ref{reduction1 2DKK}), is satisfied by several choices of the kernel $R_0$ (\ref{Kernel FP}).

{\bf{The condition I of reduction (\ref{reduction 2DKK}).}}
It is clear from (\ref{reduction1 2DKK}) that condition (\ref{reduction 2DKK}) of reduction for the simplest case $R=\pi f_1  g_1$ of $N=1$ of one term in the sum
(\ref{Kernel FP}) is satisfied if $\alpha_1$ and $\beta_1$ proportional to each other: $\alpha_1(x,y,t)=c_1\beta_{1}(x,y,t)$. Now let us prove that under the same interrelations between $\alpha_k$ and $\beta_{k}$ , with some constants $c_k$, for all $k$,
\begin{equation}\label{CRed1 ab}
\alpha_k(x,y,t)=c_k\beta_{k}(x,y,t),\quad (k=1,\ldots,N),
\end{equation}
the condition (\ref{reduction1 2DKK}) is satisfied also for more general case of $N\neq1$ terms in the kernel    (\ref{Kernel FP}).
At first due to (\ref{CRed1 ab}) one transforms  matrix $A$ given by (\ref{matrix1_FP}),
\begin{equation}\label{matrix1_FP proof CRed}
A_{lk}=\delta_{lk}+\frac{1}{2c_k}\partial_{x}^{-1}\alpha_l \alpha_k=
\sqrt{c_l}(\delta_{lk}+\frac{1}{2\sqrt{c_k c_l}}\partial_{x}^{-1}\alpha_l \alpha_k)\frac{1}{\sqrt{c_k}}=\sqrt{c_l}\tilde{A}_{lk}\frac{1}{\sqrt{c_k}},
\end{equation}
where $\tilde{A}=\delta_{lk}+\frac{1}{2\sqrt{c_k c_l}}\partial_{x}^{-1}\alpha_l \alpha_k$ is symmetrical matrix.
One continues in this fashion transforming matrix $B$ (\ref{matrix B}),
\begin{equation}\label{matrixB_FP proof CRed}
B_{lk}=\frac{1}{c_k}\alpha_l\alpha_{k\,x}-\frac{1}{c_k}\alpha_{l\,x}\alpha_{k}=
\sqrt{c_l}(\frac{1}{\sqrt{c_k c_l}}\alpha_l\alpha_{k\,x}-\frac{1}{\sqrt{c_k c_l}}\alpha_{l\,x}\alpha_{k})\frac{1}{\sqrt{c_k}}=
\sqrt{c_l}\tilde{B}_{lk}\frac{1}{\sqrt{c_k}},
\end{equation}
where $\tilde{B}=\frac{1}{\sqrt{c_k c_l}}(\alpha_l\alpha_{k\,x}-\alpha_{l\,x}\alpha_{k})$ is antisymmetrical matrix.
Consequently under conditions (\ref{CRed1 ab}) imposed on functional parameters  the reduction condition (\ref{reduction 2DKK}) or (\ref{reduction1 2DKK}) for the 2DKK equation  due to (\ref{matrix1_FP proof CRed}) and (\ref{matrixB_FP proof CRed}) transforms to the form,
\begin{equation}
\frac{\partial}{\partial x}Tr (\tilde{B} \tilde{A}^{-1})=0,
\end{equation}
which satisfies for every $x,y,t$ due to the facts that $\tilde{A}^{-1}$ -- symmetrical as matrix inverse to symmetrical matrix $\tilde A$ and $\tilde B$ -- antisymmetrical matrix.

As a result of performed consideration it is shown that the condition of reduction (\ref{reduction 2DKK}) or (\ref{reduction1 2DKK}) for the 2DKK equation is fulfilled for the interrelations between of functional parameters $\alpha_k$ and $\beta_k$ of the type  (\ref{CRed1 ab}).  From definitions of the $\alpha_k, \beta_k$ (\ref{alpha_FP,beta_FP}) follow due to  (\ref{CRed1 ab}) the interrelations between functional parameters $f_k$ and $g_k$ in spectral representation,
\begin{equation}\label{CRed1 fg}
f_k(\mu,\overline{\mu})=c_k g_{k}(-\mu,-\overline{\mu}), \quad (k=1,...,N).
\end{equation}
 So in considered relatively simple but important and interesting case, which {\bf{will be named as condition I of reduction (\ref{reduction 2DKK})}},
 the kernel $R_0$ satisfying to the condition of reduction (\ref{reduction 2DKK}) or (\ref{reduction1 2DKK}) has due to (\ref{Kernel FP}) and (\ref{CRed1 fg}) the form,
\begin{equation}\label{Kernel FP1 1to1}
R_{0}(\mu,\overline{\mu},\lambda,\overline{\lambda})=
\pi \sum\limits_{k=1}^{N} c_k^{-1}f_k(\mu,\overline{\mu})f_k(-\lambda,-\overline{\lambda}).
\end{equation}

To the first condition of reality from (\ref{RC1})  one can satisfy by imposing on each term of the sum (\ref{Kernel FP1 1to1}) the following restrictions,
\begin{equation}\label{kernel RC1&2 1to1}
c_k^{-1}f_k(\mu,\overline{\mu})f_k(-\lambda,-\overline{\lambda})=
\overline{c}_k^{-1}\;\overline{f_k(-\overline{\mu},-\mu)}\;\overline{f_k(\overline{\lambda},\lambda)}.
\end{equation}
The meaning of such kind of condition is very simple: each term of the sum (\ref{Kernel FP1 1to1}) after applying reality condition does not change.
By separating variables,
\begin{equation}\label{RC1 1to1 separate}
\frac{f_k(\mu,\overline{\mu})}{\overline{f_k(-\overline{\mu},-\mu)}}=
\frac{\overline{c}_k^{-1}\;\overline{f_k(\overline{\lambda},\lambda)}}
{c_k^{-1}\;f_k(-\lambda,-\overline{\lambda})}=v_k,
\end{equation}
with some complex constants $v_k$, one obtain the following restrictions on the functions $f_k(\mu,\overline{\mu})$,
\begin{equation}\label{RC1 1to1 f}
f_k(\mu,\overline{\mu})=v_k\overline{f_k(-\overline{\mu},-\mu)},\quad
\frac{v_k}{c_k}=\overline{\left(\frac{v_k}{c_k}\right)},\;
v_k^{2}=\frac{c_k}{\overline{c}_k},\; |v_k|=1.
\end{equation}
Second condition of reality from (\ref{RC1}) leads effectively to exactly the same relations as in (\ref{RC1 1to1 f}) and will not be considered here.

Due to definitions (\ref{alpha_FP,beta_FP}) and (\ref{CRed1 fg}),(\ref{RC1 1to1 f}) one  derive the following interrelations between different functional parameters,
\begin{equation}\label{C1alpha,beta CRed1 1to1}
\alpha_k: = v_k\int\int\limits_C {\overline{f_k(-\overline{\mu},-\mu)}}e^{F(\mu)}d\mu\wedge d\overline{\mu}=-v_k\overline{\alpha_k},\quad
\beta_k: = c_k^{-1}\int\int\limits_C f_k(-\lambda,-\overline{\lambda})e^{-F(\lambda)}d\lambda \wedge d\overline{\lambda}=c_k^{-1}{\alpha_k}.
\end{equation}
So due to (\ref{C1alpha,beta CRed1 1to1}) the sets of functional parameters are characterized by the following properties,
\begin{equation}\label{set of alpha,beta CRed1 1to1}
(\alpha_1,\ldots,\alpha_{N}):=(-v_1\overline{\alpha_1},\ldots, -v_N\overline{\alpha_N}),\quad
(\beta_1,\ldots,\beta_{N}):=(c_1^{-1}\alpha_1,\ldots, c_N^{-1}\alpha_{N})
\end{equation}
i.e. both sets $\alpha_n$ and $\beta_n$ express through $N$ independent complex functional parameters $(\alpha_1(x,y,t),\ldots, \alpha_N(x,y,t))$ satisfying to the first relation from (\ref{C1alpha,beta CRed1 1to1}).

General determinant formula (\ref{Solution_general_formula_FP}) with matrix $A$ (\ref{matrix1_FP}) corresponding to  kernel $R_0$
(\ref{Kernel FP1 1to1}) of  $\overline{\partial}$-problem (\ref{dibar problem}) gives the class of exact solutions $u$ with functional parameters for  2DKK equation (\ref{2DKK}).
By construction due to (\ref{set of alpha,beta CRed1 1to1}) these solutions depend on $N$ functional parameters $(\alpha_1,\ldots,\alpha_N)$.
In the simplest case $N=1$, $\beta_1:=c_1^{-1}\alpha_1$ and due to (\ref{C1alpha,beta CRed1 1to1})  determinant of matrix $A$  (\ref{matrix1_FP}) has the form,
\begin{eqnarray}\label{determinant CRed1 1to1 N1}
\det{A}=\big(1+\frac{1}{2c_1}\partial_{x}^{-1}\alpha_1^2\big)=
\big(1-\frac{v_1}{2c_1}\partial_{x}^{-1}|\alpha_1|^2\big).
\end{eqnarray}
The corresponding solution $u$ with functional parameters of  2DKK equation due to (\ref{Solution_general_formula_FP}) and (\ref{determinant CRed1 1to1 N1}) is given by expression,
\begin{equation}\label{C1Solution CRed1 1to1 N1}
u(x,y,t)=-\frac{3v_1}{2c_1\det{A}^2}\big(\det{A}\cdot|\alpha_1|_x^2+
\frac{v_1}{2c_1}|\alpha_1|^4\big),
\end{equation}
where $v_1/c_1$ due to (\ref{RC1 1to1 f}) is some real parameter.

In the case $N=2$ $(\beta_1,\beta_2):=(c_1^{-1}\alpha_1, c_2^{-1}\alpha_{2})$ and due to (\ref{C1alpha,beta CRed1 1to1})  determinant of matrix $A$  (\ref{matrix1_FP}) has the form,
\begin{eqnarray}\label{determinant CRed1 1to1 N2}
&\det{A}=\big(1+\frac{1}{2c_1}\partial_{x}^{-1}\alpha_1^2\big)\big(1+\frac{1}{2c_2}\partial_{x}^{-1}\alpha_2^2\big)-
\frac{1}{4c_1 c_2} \big(\partial_{x}^{-1}\alpha_1\alpha_2\big)^2=\nonumber\\
&=\big(1-\frac{v_1}{2c_1}\partial_{x}^{-1}|\alpha_1|^2\big)\big(1-\frac{v_2}{2c_2}\partial_{x}^{-1}|\alpha_2|^2\big)-\frac{v_1 v_2}{4 c_1 c_2}
\Big|\partial_{x}^{-1}\alpha_1\alpha_2\Big|^2,
\end{eqnarray}
where $v_k/c_k,\, (k=1,2)$ due to (\ref{RC1 1to1 f}) are some real parameters.
The corresponding solution $u$ is calculated with help of reconstruction formula (\ref{Solution_general_formula_FP}).
It is interesting to note that  considered simplest solutions $u(x,y,t)$ of 2DKK equation with functional parameters for cases $N=1,2$  are nonsingular for negative values of constants $\frac{v_k}{c_k}<0,\, (k=1,2)$; for $N=2$ it follows from (\ref{determinant CRed1 1to1 N2}) particularly due to Cauchy-Bunyakovskii inequality $(\partial_{x}^{-1}|\alpha_1|^2)(\partial_{x}^{-1}|\alpha_2|^2)
\geq|\partial_{x}^{-1}\alpha_1\alpha_2|^2$.

In particular case of kernel $R_0$ (\ref{Kernel FP1 1to1}) of delta-functional   type with $f_k(\mu,\overline{\mu})=A_k\delta(\mu-i\mu_{k0})$,
$(k=1,\ldots,N)$, which satisfy to  conditions (\ref{CRed1 fg}) and (\ref{RC1 1to1 f}), due to definitions (\ref{alpha_FP,beta_FP}) functional parameters $\alpha_k$ have the following forms,
\begin{equation}\label{C1CRed1 1to1 N1 DeltaFunction}
\alpha_{k}=-2i A_k e^{F(i\mu_{k0})},
\end{equation}
where in accordance with (\ref{RC1 1to1 f}) $A_k=v_k \overline {A}_k$ and $\mu_{k0}$ are some real parameters. Such kernel leads to corresponding exact multiline soliton solutions.

In the simplest case of $N=1$ from (\ref{C1Solution CRed1 1to1 N1}), (\ref{C1CRed1 1to1 N1 DeltaFunction}), under the condition $\frac{v_1 |A_1|^2}{c_1 \mu_{10}}=e^{2\varphi_0}>0$, one obtains the exact nonsingular one line soliton solution of the 2DKK equation,
\begin{equation}\label{Solution CRed1 1to1 N1 DeltaFunction}
    u(x,y,t) = \frac{3 \mu_{10}^2}{\cosh^2(\mu_{10}x-\mu_{10}^3 y+9\mu_{10}^5t-\varphi_0)}.
\end{equation}
This one line soliton solution was derived earlier in paper of first author \cite{DubrLisit}. In the case of $N=2$ from (\ref{Solution_general_formula_FP}), (\ref{determinant CRed1 1to1 N2}) and (\ref{C1CRed1 1to1 N1 DeltaFunction})  can be easily calculated  exact two line soliton solution of the 2DKK equation.

To conditions of reality (\ref{RC1}) of the solution of  2DKK equation (\ref{2DKK}) one can satisfy more nontrivially by imposing another restrictions on functional parameters $f_k(\mu,\overline{\mu}),\; g_{k}(\lambda,\overline{\lambda})$ or $\alpha_k(x,y,t),\;\beta_k(x,y,t)$.
For this purpose the terms in the sum (\ref{Kernel FP}) or (\ref{Kernel FP1 1to1})
for the kernel $R_0$ can be grouped  by pairs.
The kernel of the $\overline{\partial}$-problem for which condition of reduction (\ref{reduction 2DKK}) or (\ref{reduction1 2DKK}) fulfils,
has due to (\ref{Kernel FP1 1to1}) the form,
\begin{equation}\label{Kernel FP1 pairs1}
R_{0}(\mu,\overline{\mu},\lambda,\overline{\lambda})=
\pi\sum\limits_{k=1}^{2N}c_k^{-1}f_k(\mu,\overline{\mu})f_k(-\lambda,-\overline{\lambda})=
\pi\sum\limits_{k=1}^{N}\Big[c_k^{-1} p_k(\mu,\overline{\mu})
p_k(-\lambda,-\overline{\lambda})+\tilde{c}_k^{-1}
\tilde{p}_k(\mu,\overline{\mu})\tilde{p}_k(-\lambda,-\overline{\lambda})\Big],
\end{equation}
where $(f_1,\ldots,f_{2N})=(p_1(\mu,\overline{\mu}),\ldots, p_N(\mu,\overline{\mu});\tilde{p}_1(\mu,\overline{\mu}),\ldots, \tilde{p}_N(\mu,\overline{\mu}))$.
To the first and the second conditions of reality from (\ref{RC1})  one can satisfy by imposing on each pair of terms in the sum (\ref{Kernel FP1 pairs1}) the following restrictions,
\begin{equation}\label{RC1 1to2}
c_k^{-1} p_k(\mu,\overline{\mu})p_k(-\lambda,-\overline{\lambda})=
\overline{\tilde{c}}_k^{-1}\;\overline{\tilde{p}_k(-\overline{\mu},-\mu)}\;
\overline{\tilde{p}_k(\overline{\lambda},\lambda)},
\end{equation}
i. e. the first term in square bracket of (\ref{Kernel FP1 pairs1}) goes under considered reality conditions to second one.
By separating variables in last expression (\ref{RC1 1to2}),
\begin{equation}\label{RC1 1to2_separate}
\frac{p_k(\mu,\overline{\mu})}{\overline{\tilde{p}_k(-\overline{\mu},-\mu)}}=
\frac{\overline{\tilde{c}}_k^{-1}\overline{\tilde{p}_k(\overline{\lambda},\lambda)}}{c_k^{-1} p_k(-\lambda,-\overline{\lambda})}=\overline{v}_k^{-1},
\end{equation}
with some complex constants $v_k$, one obtain the interrelations between the functional parameters $\tilde{p}_k(\mu,\overline{\mu})$ and $p_k(\mu,\overline{\mu})$,
\begin{equation}\label{RC1 1to2_p}
\tilde{p}_k(\mu,\overline{\mu})=v_k \overline{p_k(-\overline{\mu},-\mu)},\quad
\tilde{c}_k=v_k^{2}\overline{c}_k,\quad (k=1,...,N).
\end{equation}

So the kernel which satisfies to the first and the second conditions of reality from (\ref{RC1}) and condition of reduction (\ref{reduction 2DKK}) or (\ref{reduction1 2DKK})  due to (\ref{CRed1 fg}), (\ref{Kernel FP1 pairs1}), (\ref{RC1 1to2_p}) has the form,
\begin{equation}\label{kernel_satify CRed&CR 1to2}
R_{0}(\mu,\overline{\mu},\lambda,\overline{\lambda})=
\pi\sum\limits_{k=1}^{N}\Big[c_k^{-1}p_k(\mu,\overline{\mu})p_k(-\lambda,-\overline{\lambda})+
\overline{c}_k^{-1}\overline{p_k(-\overline{\mu},-\mu)}\;\overline{p_k(\overline{\lambda},\lambda)}\Big],
\end{equation}
and due to (\ref{Kernel FP1 pairs1}),  (\ref{kernel_satify CRed&CR 1to2}) one can choose  the following convenient sets $f$ and $g$ of functions $f_n$, $g_n$, $n=1\ldots 2N$,
\begin{equation}\label{set of f CRed&CR 1to2}
f:=(f_1,\ldots,f_{2N})=(p_1(\mu,\overline{\mu}),\ldots, p_N(\mu,\overline{\mu});\overline{p_1(-\overline{\mu},-\mu)},\ldots,\overline{p_N(-\overline{\mu},-\mu)}),
\end{equation}
\begin{equation}\label{set of g CRed&CR 1to2}
g:=(g_1,\ldots,g_{2N})=(c_1^{-1}p_1(-\lambda,-\overline{\lambda}),\ldots, c_N^{-1}p_N(-\lambda,-\overline{\lambda});
\overline{c}_1^{-1}\overline{p_1(\overline{\lambda},\lambda)},\ldots,\overline{c}_N^{-1}\overline{p_N(\overline{\lambda},\lambda)}).
\end{equation}
Due to definitions (\ref{alpha_FP,beta_FP}) and (\ref{set of f CRed&CR 1to2}),(\ref{set of g CRed&CR 1to2}) one  derive the following
interrelations between different functional parameters in coordinate representation,
\begin{equation}\label{beta CRed1 1to2}
\beta_k: = c_k^{-1}\int\int\limits_C p_k(-\lambda,-\overline{\lambda})e^{-F(\lambda)}d\lambda \wedge d\overline{\lambda}=c_k^{-1}{\alpha_k},\quad
\beta_{k+N}: =\overline{c}_k^{-1} \int\int\limits_C \overline{p_k(\overline{\lambda},\lambda)}e^{-F(\lambda)}d\lambda \wedge d\overline{\lambda}=-\overline{c}_k^{-1}\overline{\alpha_k},
\end{equation}
\begin{equation}\label{alpha,beta1 CRed1 1to2}
\alpha_{k+N}: = \int\int\limits_C \overline{p_k(-\overline{\mu},-\mu)}e^{F(\mu)}d\mu\wedge d\overline{\mu}=-\overline{\alpha_k},
\end{equation}
in formulas (\ref{beta CRed1 1to2}) and (\ref{alpha,beta1 CRed1 1to2}) index $k$ takes the values: $k=1,...,N$.
So due to (\ref{beta CRed1 1to2}), (\ref{alpha,beta1 CRed1 1to2}) the sets of functional parameters $\alpha_n$ and $\beta_n$ have the following properties,
\begin{equation}\label{set of alpha CRed1 1to2}
(\alpha_1,\ldots,\alpha_{2N}):=(\alpha_1,\ldots,\alpha_N; -\overline{\alpha_1},\ldots,-\overline{\alpha_N})
\end{equation}
\begin{equation}\label{set of beta CRed1 1to2}
(\beta_1,\ldots,\beta_{2N}):=(c_1^{-1}\alpha_1,\ldots, c_N^{-1}\alpha_{N}; -\overline{c}_1^{-1}\overline{\alpha}_1,\ldots, -\overline{c}_N^{-1}\overline{\alpha}_N),
\end{equation}
i.e. both sets express through $N$ independent complex functional parameters $(\alpha_1,\ldots, \alpha_N)$.

General determinant formula (\ref{Solution_general_formula_FP}) with matrix $A$ (\ref{matrix1_FP}) corresponding to  kernel $R_0$
(\ref{kernel_satify CRed&CR 1to2}) of  $\overline{\partial}$-problem (\ref{dibar problem}) gives another class of exact solutions $u$ with functional parameters of  2DKK equation (\ref{2DKK}).
By construction due to (\ref{set of alpha CRed1 1to2}), (\ref{set of beta CRed1 1to2}) these solutions depend on $N$  functional parameters $(\alpha_1,\ldots,\alpha_N)$ given by expressions (\ref{alpha_FP,beta_FP}).
In the simplest case $N=1$ $(\alpha_1,\alpha_{2}):=(\alpha_1,-\overline{\alpha_1})$,
$(\beta_1,\beta_2):=(c_1^{-1}\alpha_1,-\overline{c}_1^{-1}\overline{\alpha}_1)$ and due to (\ref{beta CRed1 1to2}),
(\ref{alpha,beta1 CRed1 1to2}) the determinant of matrix $A$  (\ref{matrix1_FP}) is given by expression:
\begin{equation}\label{determinant CRed1 1to2 N1}
\det{A}=\big(1+\frac{1}{2c_1}\partial_{x}^{-1}\alpha_1^2\big)
\big(1+\frac{1}{2\overline{c}_1}\partial_{x}^{-1}\overline{\alpha_1}^2\big)-
\frac{1}{4c_1 \overline{c}_1} \big(\partial_{x}^{-1}\alpha_1\overline{\alpha_1}\big)^2
=\big|1+\frac{1}{2c_1}\partial_{x}^{-1}\alpha_1^2\big|^2-
\frac{1}{4|c_1|^2}\Big(\partial_{x}^{-1}|\alpha_1|^2\Big)^2.
\end{equation}
Corresponding solution $u$ is calculated with help of reconstruction formula (\ref{Solution_general_formula_FP}) and has the form,
\begin{eqnarray}\label{Sol2DKK1}
u=\frac{3}{(\det{A})^2}
\Bigg(\det{A}\Big(\frac{\alpha_1\alpha_{1x}}{c}+\frac{\overline{\alpha}_1\overline{\alpha}_{1x}}{\overline{c}}
+\frac{1}{2|c|^2}\Big(\alpha_1\alpha_{1x}\partial_x^{-1}\overline{\alpha}_1^2+\overline{\alpha}_1\overline{\alpha}_{1x}
\partial_x^{-1}\alpha_1^2-|\alpha_1|_x^2\partial_x^{-1}|\alpha_1|^2
\Big)\Big)-\nonumber\\
-\Big(\frac{\alpha_1^2}{2c}+\frac{\overline{\alpha}_1^2}{2\overline{c}}
+\frac{1}{4|c|^2}\Big(\alpha_1^2\partial_x^{-1}\overline{\alpha}_1^2+\overline{\alpha}_1^2
\partial_x^{-1}\alpha_1^2-2|\alpha_1|^2\partial_x^{-1}|\alpha_1|^2
\Big)\Big)^2\Bigg),
\end{eqnarray}
and due to (\ref{determinant CRed1 1to2 N1}) evidently is singular.

In the case of  kernel $R_0$  (\ref{kernel_satify CRed&CR 1to2}) of  delta-functional type with $p_k(\mu,\overline{\mu})=A_k\delta(\mu-\mu_k)$, $(k=1,\ldots,N)$, which satisfies to the conditions
(\ref{CRed1 fg}) and (\ref{RC1 1to2_p}), due to the definitions (\ref{alpha_FP,beta_FP}) functional parameters $\alpha_{k}$ have the following form,
\begin{equation}\label{CRed1 1to2 N1 DeltaFunction}
\alpha_{k}=-2i A_k e^{F(\mu_{k})}, \quad (k=1,\ldots,N).
\end{equation}
Such kernel leads to corresponding exact multiline soliton solutions. In the simplest case of $N=1$  from
 (\ref{determinant CRed1 1to2 N1}) and (\ref{Sol2DKK1}) due to (\ref{CRed1 1to2 N1 DeltaFunction}) follows the exact one line soliton solution of  2DKK equation,
\begin{equation}
u=\frac{12}{(\det{A})^2}
\Bigg(\frac{-iA_1^2\mu_1}{c_1}e^{2F(\mu_{1})}+\frac{i\overline{A}_1^2\overline{\mu}_1}{\overline{c}_1}e^{-2F(\overline{\mu}_{1})}+
\frac{i|A|^4\mu_{1R}^2}{|c_1|^2\mu_{1I}^2}
\Big(\frac{A_1^2\overline{\mu}_1}{c_1\mu_1^2}e^{2F(\mu_{1})}
-\frac{\overline{A}_1^2\mu_1}{\overline{c}_1\overline{\mu}_1^2}e^{-2F(\overline{\mu}_{1})}+
\frac{8i\mu_{1I}^2}{|\mu_1|^2}\Big)e^{2F(\mu_{1})-2F(\overline{\mu}_{1})}\Bigg),
\end{equation}
where determinant of the matrix $A$ has form,
\begin{equation}\label{detAl}
\det{A}=1+\frac{iA_1^2}{c_1\mu_1}e^{2F(\mu_{1})}-
\frac{i\overline{A}_1^2}{\overline{c}_1\overline{\mu}_1}e^{2\overline{F(\mu_{1})}}-\frac{|A|^4\mu_{1R}^2}{|c_1|^2|\mu_1|^2\mu_{1I}^2}
e^{2F(\mu_{1})+2\overline{F(\mu_{1})}}=\big|1+\frac{iA_1^2}{c_1\mu_1}e^{2F(\mu_{1})}\big|^2-
\frac{|A|^4}{|c_1|^2\mu_{1I}^2}
e^{2F(\mu_{1})+2\overline{F(\mu_{1})}}.
\end{equation}
Due to the  expression (\ref{detAl}) for $\det{A}$ last calculated  solution evidently is singular.

{\bf{The condition II of reduction (\ref{reduction 2DKK}).}}
The condition of reduction (\ref{reduction 2DKK}) or (\ref{reduction1 2DKK}) can be satisfied by another
restriction on functional parameters. One  groups for this the terms in the kernel $R_0$ by pairs,
\begin{equation}\label{Kernel FP1 pairs}
R_{0}(\mu,\overline{\mu},\lambda,\overline{\lambda})=
\pi\sum\limits_{k=1}^{2N}f_k(\mu,\overline{\mu})g_k(\lambda,\overline{\lambda})=
\pi\sum\limits_{k=1}^{N}\Big[p_k(\mu,\overline{\mu})q_k(\lambda,\overline{\lambda})+
\tilde{p}_k(\mu,\overline{\mu})\tilde{q}_k(\lambda,\overline{\lambda})\Big].
\end{equation}
One define the following sets $f$ and $g$ of functions $f_n$, $g_n$, $n=1\ldots 2N$,
\begin{eqnarray}\label{CRed1_sets fg KK}
f:=(f_1,\ldots,f_{2N})=(p_1(\mu,\overline{\mu}),\ldots, p_N(\mu,\overline{\mu});\tilde{p}_1(\mu,\overline{\mu}),\ldots, \tilde{p}_N(\mu,\overline{\mu})),\nonumber\\
g:=(g_1,\ldots,g_{2N})=(q_1(\lambda,\overline{\lambda}),\ldots, q_N(\lambda,\overline{\lambda});\tilde{q}_1(\lambda,\overline{\lambda}),\ldots, \tilde{q}_N(\lambda,\overline{\lambda})).
\end{eqnarray}
For simplicity let us rewrite the last
expression in (\ref{reduction1 2DKK}) for case $N=1$
of one pair of terms in kernel (\ref{Kernel FP1 pairs}),
\begin{equation}\label{CRed1 matrix}
\frac{\partial}{\partial x}Tr (B A^{-1})=\frac{\partial}{\partial x}\frac{1}{\det{A}}\Big(B_{11}A_{22}-B_{12}A_{21}-
B_{21}A_{12}+B_{22}A_{11}\Big)=0,
\end{equation}
where $A_{ij}$, $B_{ij}$ are elements of matrixes $A$ (\ref{matrix1_FP}) and $B$ (\ref{matrix B}).
Substituting $A_{ij}$, $B_{ij}$ from (\ref{matrix1_FP}) and (\ref{matrix B}) into (\ref{CRed1 matrix}) one obtains, for one pair of terms in kernel (\ref{Kernel FP1 pairs}), from the equality $B_{11}A_{22}-B_{12}A_{21}-B_{21}A_{12}+B_{22}A_{11}=0$ new condition of reduction in terms of functional parameters,
\begin{eqnarray}\label{CRed2 AlphaBeta}
&\alpha_1\beta_{1x}-\alpha_{1x}\beta_{1}+\alpha_2\beta_{2x}-\alpha_{2x}\beta_{2}=
\frac{1}{2}(\alpha_1\beta_{2x}-\alpha_{1x}\beta_{2})\partial_{x}^{-1}\alpha_2\beta_{1}+
\frac{1}{2}(\alpha_2\beta_{1x}-\alpha_{2x}\beta_{1})\partial_{x}^{-1}\alpha_1\beta_{2}-\nonumber\\
&-\frac{1}{2}(\alpha_1\beta_{1x}-\alpha_{1x}\beta_{1})\partial_{x}^{-1}\alpha_2\beta_{2}-
\frac{1}{2}(\alpha_2\beta_{2x}-\alpha_{2x}\beta_{2})\partial_{x}^{-1}\alpha_1\beta_{1}.
\end{eqnarray}
It is easy to check that choice $\alpha_2=c\beta_1$ and $\beta_2=c^{-1}\alpha_1$, or due to (\ref{alpha_FP,beta_FP}) corresponding choice in terms of functional parameters in spectral representation, $f_2(\mu,\overline{\mu})=c g_1(-\mu,-\overline{\mu})$ and $g_2(\mu,\overline{\mu})=c^{-1} f_1(-\mu,-\overline{\mu})$, satisfies to condition  (\ref{CRed2 AlphaBeta}). These conditions can be generalized for case of kernel consisting of $N$ pairs in (\ref{Kernel FP1 pairs}) with following identification of multipliers,
\begin{equation}\label{CRed2}
\tilde{p}_k(\mu,\overline{\mu})=c_k q_k(-\mu,-\overline{\mu}),\quad
\tilde{q}_k(\mu,\overline{\mu})=c^{-1}_k p_k(-\mu,-\overline{\mu}), \quad (k=1,...,N).
\end{equation}
So to the condition of reduction (\ref{reduction 2DKK}) or (\ref{reduction1 2DKK}) is satisfied due to (\ref{CRed2}) following kernel $R_0$ (\ref{Kernel FP1 pairs}) of the $\overline{\partial}$-problem (\ref{dibar problem}),
\begin{equation}\label{Kernel FP1 pairs CRed2}
R_{0}(\mu,\overline{\mu},\lambda,\overline{\lambda})=
\pi\sum\limits_{k=1}^{2N}f_k(\mu,\overline{\mu})g_k(\lambda,\overline{\lambda})=\pi\sum\limits_{k=1}^{N}\Big[p_k(\mu,\overline{\mu})
q_k(\lambda,\overline{\lambda})+q_k(-\mu,-\overline{\mu})p_k(-\lambda,-\overline{\lambda})\Big].
\end{equation}
Formulated conditions (\ref{CRed2}) will be named as {\bf{condition II of reduction (\ref{reduction 2DKK}).}}

The first condition of reality from (\ref{RC1}) and condition of reduction (\ref{reduction 2DKK}) or (\ref{reduction1 2DKK}) are satisfied simultaneously by imposing on each pair of terms in the sum (\ref{Kernel FP1 pairs CRed2}) the following restriction,
\begin{eqnarray}\label{kernel RC1 CRed2}
p_k(\mu,\overline{\mu})q_k(\lambda,\overline{\lambda})+
q_k(-\mu,-\overline{\mu})p_k(-\lambda,-\overline{\lambda})=\nonumber\\
=\overline{p_k(-\overline{\mu},-\mu)}\;\overline{q_k(-\overline{\lambda},-\lambda)}+
\overline{q_k(\overline{\mu},\mu)}\;\overline{p_k(\overline{\lambda},\lambda)}.
\end{eqnarray}
Due to (\ref{kernel RC1 CRed2}) two cases are possible,
\begin{equation}\label{RC1 CRed2 1to1}
A.\quad p_k(\mu,\overline{\mu})q_k(\lambda,\overline{\lambda})=
\overline{p_k(-\overline{\mu},-\mu)}\;\overline{q_k(-\overline{\lambda},-\lambda)},
\end{equation}
\begin{equation}\label{RC1 CRed2 1to2}
B.\quad p_k(\mu,\overline{\mu})q_k(\lambda,\overline{\lambda})=
\overline{q_k(\overline{\mu},\mu)}\;\overline{p_k(\overline{\lambda},\lambda)}.
\end{equation}

In the case $A.$ by separating variables,
\begin{equation}\label{RC1 CRed2 1to1 separate}
\frac{p_k(\mu,\overline{\mu})}{\overline{p_k(-\overline{\mu},-\mu)}}=
\frac{\overline{q_k(-\overline{\lambda},-\lambda)}}{q_k(\lambda,\overline{\lambda})}=v_k,
\end{equation}
with some complex constants $v_k, (k=1,...,N)$,
one obtain the following restrictions on the functions $p_k(\mu,\overline{\mu})$ and $q_k(\lambda,\overline{\lambda})$,
\begin{equation}\label{RC1 CRed2 1to1 p}
p_k(\mu,\overline{\mu})=v_k\overline{p_k(-\overline{\mu},-\mu)},\quad
q_k(\lambda,\overline{\lambda})=v^{-1}_k\overline{q_k(-\overline{\lambda},-\lambda)}.
\end{equation}

In the case $B.$ by separating variables,
\begin{equation}\label{RC1 CRed2 1to2_separate}
\frac{p_k(\mu,\overline{\mu})}{\overline{q_k(\overline{\mu},\mu)}}=
\frac{\overline{p_k(\overline{\lambda},\lambda)}}{q_k(\lambda,\overline{\lambda})}=v_k,
\end{equation}
with some another constants $v_k, (k=1,...,N)$, one obtain  another restrictions on the functions $p_k(\mu,\overline{\mu})$ and $q_k(\lambda,\overline{\lambda})$,
\begin{equation}\label{RC1 CRed2 1to2_p}
q_k(\mu,\overline{\mu})=v^{-1}_k \overline{p_k(\overline{\mu},\mu)},\quad
\overline{v_k}=v_k.
\end{equation}

One can show that second condition of reality from (\ref{RC1})
leads to the same restrictions on the kernel $R_0$
as obtained above; i. e. only the cases A.
(\ref{RC1 CRed2 1to1}) and B. (\ref{RC1 CRed2 1to2}) will be discussed further.

So for the case $A.$ (\ref{RC1 CRed2 1to1}) the kernel $R_0$ which satisfies simultaneously to conditions  of reduction (\ref{reduction 2DKK}) or (\ref{reduction1 2DKK})  and reality (\ref{RC1})  due to (\ref{CRed2}) and (\ref{Kernel FP1 pairs CRed2}) takes the form,
\begin{equation}\label{kernel_satify_CRed2&CR 1to1}
R_{0}(\mu,\overline{\mu},\lambda,\overline{\lambda})=
\pi\sum\limits_{k=1}^{2N}f_k(\mu,\overline{\mu})g_k(\lambda,\overline{\lambda})=\pi\sum\limits_{k=1}^{N}\Big[p_k(\mu,\overline{\mu})
q_k(\lambda,\overline{\lambda})+q_k(-\mu,-\overline{\mu})p_k(-\lambda,-\overline{\lambda})\Big],
\end{equation}
where functions $p_k$ and $q_k$ are satisfy to conditions (\ref{RC1 CRed2 1to1 p}).

For the kernel (\ref{kernel_satify_CRed2&CR 1to1}) one  choose  the following convenient sets $f$ and $g$ of functions $f_n$, $g_n$, $n=1\ldots 2N$,
\begin{equation}\label{set of f CRed2&CR 1to1}
f:=(f_1,\ldots,f_{2N})=(p_1(\mu,\overline{\mu}),\ldots, p_N(\mu,\overline{\mu}); q_1(-\mu,-\overline{\mu}),\ldots, q_N(-\mu,-\overline{\mu})),
\end{equation}
\begin{equation}\label{set of g CRed2&CR 1to1}
g:=(g_1,\ldots,g_{2N})=(q_1(\lambda,\overline{\lambda}),\ldots, q_N(\lambda,\overline{\lambda});
 p_1(-\lambda,-\overline{\lambda}),\ldots, p_N(-\lambda,-\overline{\lambda})).
\end{equation}
Due to definitions (\ref{alpha_FP,beta_FP}) and (\ref{RC1 CRed2 1to1 p}), (\ref{set of f CRed2&CR 1to1}), (\ref{set of g CRed2&CR 1to1}) one  derive the following
interrelations between different functional parameters,
\begin{equation}\label{alpha,beta CRed2 1to1}
\alpha_{k}: =v_k \int\int\limits_C \overline{p_k(-\overline{\mu},-\mu)}e^{F(\mu)}d\mu\wedge d\overline{\mu}=-v_k\overline{\alpha_k},\quad
\beta_{k}: =v^{-1}_k \int\int\limits_C \overline{q_k(-\overline{\lambda},-\lambda)}e^{-F(\lambda)}
d\lambda \wedge d\overline{\lambda}=-v^{-1}_k\overline{\beta_k},
\end{equation}
\begin{equation}\label{alpha1,beta1 CRed2 1to1}
\alpha_{k+N}: = \int\int\limits_C q_k(-\mu,-\overline{\mu})e^{F(\mu)}d\mu \wedge d\overline{\mu}= \beta_k,\quad
\beta_{k+N}: = \int\int\limits_C  p_k(-\lambda,-\overline{\lambda})e^{-F(\lambda)}d\lambda \wedge d\overline{\lambda}=\alpha_k,
\end{equation}
where in formulas (\ref{alpha,beta CRed2 1to1}) and (\ref{alpha1,beta1 CRed2 1to1}) index $k$ takes the values: $k=1,...,N$.
So by the use (\ref{alpha,beta CRed2 1to1}),(\ref{alpha1,beta1 CRed2 1to1}) one concludes that the sets of functional parameters have the following structure,
\begin{equation}\label{set of alpha CRed2 1to1}
(\alpha_1,\ldots,\alpha_{2N}):=(\alpha_1,\ldots,\alpha_N;  \beta_1,\ldots, \beta_N)
\end{equation}
\begin{equation}\label{set of beta CRed2 1to1}
(\beta_1,\ldots,\beta_{2N}):=(\beta_1,\ldots,\beta_N; \alpha_1,\ldots, \alpha_N)
\end{equation}
i.e. both sets are expressed through $2N$ independent functional parameters $(\alpha_1,\ldots, \alpha_N)$ and $(\beta_1,\ldots, \beta_N)$ with properties (\ref{alpha,beta CRed2 1to1}).

General determinant formula (\ref{Solution_general_formula_FP}) with matrix $A$ (\ref{matrix1_FP}) corresponding to  kernel $R_0$
(\ref{kernel_satify_CRed2&CR 1to1}) of  $\overline{\partial}$-problem (\ref{dibar problem}) gives the class of exact solutions $u$ with functional parameters of  2DKK equation (\ref{2DKK}).
By construction due to (\ref{set of alpha CRed2 1to1}), (\ref{set of beta CRed2 1to1}) these solutions depend on $2N$ functional parameters $(\alpha_1,\ldots,\alpha_N)$ and $(\beta_1,\ldots, \beta_N)$ with properties  (\ref{alpha,beta CRed2 1to1}).

In the simplest case $N=1$ $(\alpha_1,\alpha_{2}):=(\alpha_1, \beta_1)$,
$(\beta_1,\beta_2):=(\beta_1, \alpha_1)$ the determinant of matrix $A$ due to (\ref{matrix1_FP}) is given by expression,
\begin{eqnarray}\label{determinant CRed2 1to1 N1}
\det{A}=\big(1+\frac{1}{2}\partial_{x}^{-1}\alpha_1\beta_1\big)^2-
\frac{1}{4}\partial_x^{-1}|\alpha_1|^{2}\:\partial_x^{-1}|\beta_1|^{2},
\end{eqnarray}
where due to (\ref{alpha,beta CRed2 1to1}) $\alpha_1\beta_1=\overline{\alpha_1}\;\overline{\beta_1}$.
The corresponding solution $u$ is calculated with help of reconstruction formula (\ref{Solution_general_formula_FP}) and has the form,
\begin{eqnarray}\label{SimpleSol}
u=\frac{3}{(\det{A})^2}
\Bigg(\det{A}\Big(\frac{1}{2}\alpha_1^2\beta_1^2-\frac{1}{2}|\alpha_1|^2|\beta_1|^2+
(\alpha_1\beta_1)_x(1+\frac{1}{2}\partial_x^{-1}\alpha_1\beta_1)-\frac{1}{4}|\alpha_1|_x^2\partial_x^{-1}|\beta_1|^2-
\frac{1}{4}|\beta_1|_x^2\partial_x^{-1}|\alpha_1|^2\Big)-\nonumber \\
-\Big(\alpha_1\beta_1(1+\frac{1}{2}\partial_x^{-1}\alpha_1\beta_1)-
\frac{1}{4}|\alpha_1|^2\partial_x^{-1}|\beta_1|^2-\frac{1}{4}
|\beta_1|^2\partial_x^{-1}|\alpha_1|^2\Big)^2\Bigg).
\end{eqnarray}

The calculated simplest solutions $u(x,y,t)$ (\ref{SimpleSol}) with functional parameters of 2DKK equation   for case $N=1$ may be due (\ref{determinant CRed2 1to1 N1})  singular or nonsingular, it depends on concrete choice of functional parameters.

For the case of  kernel $R_0$ (\ref{kernel_satify_CRed2&CR 1to1}) of delta-functional type,  with
$p_k(\mu,\overline{\mu})=A_k\delta(\mu-i\mu_{k0})$, $q_k(\lambda,\overline{\lambda})=B_k\delta(\lambda-i\lambda_{k0})$,
$(k=1,\ldots,N)$, which satisfies the conditions (\ref{CRed2}) and (\ref{RC1 CRed2 1to1 p}), the functional parameters have the following form,
\begin{equation}\label{CRed2 1to1 N1 DeltaFunction}
\alpha_{k}=-2i A_k e^{F(i\mu_{k0})},\quad \beta_{k}=-2i B_k e^{-F(i\lambda_{k0})}, \quad (k=1,\ldots,N),
\end{equation}
where in accordance with (\ref{RC1 CRed2 1to1 p}) $A_k=v_k \overline {A}_k$, $B_k=v^{-1}_k \overline {B}_k$ and $\mu_{k0}$, $\lambda_{k0}$ - some real parameters. Such kernel leads to corresponding exact multiline soliton solutions.
In the simplest case of $N=1$  one obtains from (\ref{determinant CRed2 1to1 N1}), (\ref{SimpleSol}) via
(\ref{CRed2 1to1 N1 DeltaFunction}) the exact one line soliton solution of the 2DKK equation,
\begin{eqnarray}\label{Solution CRed2 1to1 N1 DeltaFunction}
\det{A}=1+\frac{4a}{\mu_{10}-\lambda_{10}}e^{\varphi(x, y, t)}+\frac{a^2}{\mu_{10}\lambda_{10}}
\Bigg(\frac{\mu_{10}+\lambda_{10}}{\mu_{10}-\lambda_{10}}\Bigg)^2e^{2\varphi(x, y, t)},\nonumber\\
u(x, y, t) = \frac{12 a (\mu_{10}-\lambda_{10})e^{\varphi}}{\det{A}^2}\Bigg[\det{A}+\frac{a(\mu_{10}-\lambda_{10})e^{\varphi}}{\mu_{10}\lambda_{10}}\Bigg],
\end{eqnarray}
where $a=A_1B_1=\bar a$ - some real parameter, $\varphi(x, y, t):=F(i\mu_{10})-F(i\lambda_{10})$.
This one line soliton solution for the values of parameters, $\frac{a}{\mu_{10}-\lambda_{10}}>0,\mu_{10}\lambda_{10}>0$, is nonsingular and  was derived earlier in paper of first author \cite{DubrLisit}.

For the case $B.$ (\ref{RC1 CRed2 1to2}) the kernel $R_0$ which satisfies conditions of reduction (\ref{reduction 2DKK}) or (\ref{reduction1 2DKK}) and reality (\ref{RC1}) due to (\ref{CRed2}), (\ref{Kernel FP1 pairs CRed2}) and (\ref{RC1 CRed2 1to2_p}) has the form,
\begin{equation}\label{kernel_satify_CRed2&CR 1to2}
R_{0}(\mu,\overline{\mu},\lambda,\overline{\lambda})=
\pi\sum\limits_{k=1}^{2N}f_k(\mu,\overline{\mu})g_k(\lambda,\overline{\lambda})=
\pi\sum\limits_{k=1}^{N}\Big[v^{-1}_k p_k(\mu,\overline{\mu})
\overline{p_k(\overline{\lambda},\lambda)}+v^{-1}_k \overline{p_k(-\overline{\mu},-\mu)}
p_k(-\lambda,-\overline{\lambda})\Big].
\end{equation}
From (\ref{kernel_satify_CRed2&CR 1to2}) one choose  the following convenient sets $f$ and $g$ of functions $f_n$, $g_n$, $n=1\ldots2N$,
\begin{equation}\label{set of f CRed2&CR 1to2}
f:=(f_1,\ldots,f_{2N})=(p_1(\mu,\overline{\mu}),\ldots, p_N(\mu,\overline{\mu}); v^{-1}_1\overline{p_1(-\overline{\mu},-\mu)},\ldots, v^{-1}_N\overline{p_N(-\overline{\mu},-\mu)}),
\end{equation}
\begin{equation}\label{set of g CRed2&CR 1to2}
g:=(g_1,\ldots,g_{2N})=(v^{-1}_1 \overline{p_1(\overline{\lambda},\lambda)},\ldots, v^{-1}_N \overline{p_N(\overline{\lambda},\lambda)};
p_1(-\lambda,-\overline{\lambda}),\ldots, p_N(-\lambda,-\overline{\lambda})).
\end{equation}
Due to definitions (\ref{alpha_FP,beta_FP}) and (\ref{set of f CRed2&CR 1to2}), (\ref{set of g CRed2&CR 1to2}) one derive
the following interrelations between different functional parameters,
\begin{equation}\label{beta CRed2 1to2}
\beta_{k}: =v^{-1}_k \int\int\limits_C \overline{p_k(\overline{\lambda},\lambda)}
e^{-F(\lambda)}d\lambda \wedge d\overline{\lambda}=-v^{-1}_k\overline{\alpha_k},
\qquad k=1,\ldots N;
\end{equation}
\begin{equation}\label{alpha1,beta1 CRed2 1to2}
\alpha_{k+N}: = v^{-1}_k\int\int\limits_C \overline{p_k(-\overline{\mu},-\mu)} e^{F(\mu)}d\mu \wedge d\overline{\mu}=-v^{-1}_k\overline{\alpha}_k,\quad
\beta_{k+N}: = \int\int\limits_C p_k(-\lambda,-\overline{\lambda})e^{-F(\lambda)}d\lambda \wedge d\overline{\lambda}=\alpha_k.
\end{equation}
So using (\ref{beta CRed2 1to2}), (\ref{alpha1,beta1 CRed2 1to2}) one concludes that the  sets of functional parameters have the following structure,
\begin{equation}\label{set of alpha CRed2 1to2}
(\alpha_1,\ldots,\alpha_{2N}):=(\alpha_1,\ldots,\alpha_N;-v^{-1}_1\overline{\alpha}_1,\ldots,-v^{-1}_N\overline{\alpha}_N),
\end{equation}
\begin{equation}\label{set of beta CRed2 1to2}
(\beta_1,\ldots,\beta_{2N}):=(-v^{-1}_1\overline{\alpha_1},\ldots,-v^{-1}_N\overline{\alpha_N}; \alpha_1,\ldots, \alpha_N),
\end{equation}
i.e. both sets express through $N$ independent complex functional parameters $(\alpha_1,\ldots, \alpha_N)$.

General determinant formula (\ref{Solution_general_formula_FP}) with matrix $A$ (\ref{matrix1_FP}) corresponding to  kernel $R_0$
(\ref{kernel_satify_CRed2&CR 1to2}) of  $\overline{\partial}$-problem (\ref{dibar problem}) gives the class of exact solutions $u$ with functional parameters of the 2DKK equation (\ref{2DKK}).
By construction due to (\ref{set of alpha CRed2 1to2}), (\ref{set of beta CRed2 1to2}) these solutions depend on $N$ functional parameters $(\alpha_1,\ldots,\alpha_N)$ given by
(\ref{alpha_FP,beta_FP}).

In the simplest case $N=1$
$(\alpha_1,\alpha_{2}):=(\alpha_1,- v^{-1}_1\overline{\alpha}_1)$,
$(\beta_1,\beta_2):=(-v^{-1}_1\overline{\alpha_1}, \alpha_1)$
the determinant of $A$ due to (\ref{matrix1_FP}) is given by expression,
\begin{eqnarray}\label{Determinant CRed2 1to2 N1}
\det{A}=\big(1-\frac{1}{2v_1}\partial_{x}^{-1}|\alpha_1|^2\big)^2-
\frac{1}{4v_1^2}\Big|\partial_x^{-1}\alpha_1^{2}\Big|^2,
\end{eqnarray}
where due to (\ref{RC1 CRed2 1to2_p}) $v_1=\overline{v_1}$.
The corresponding solution $u$ is calculated with help of reconstruction formula (\ref{Solution_general_formula_FP}); for $v_1<0$, due to (\ref{Determinant CRed2 1to2 N1}) and  to Cauchy-Bunyakovskii inequality $(\partial_{x}^{-1}|\alpha_1|^2)(\partial_{x}^{-1}|\alpha_1|^2)
\geq|\partial_{x}^{-1}\alpha_1\alpha_1|^2$, this solution  is nonsingular and has the form,
\begin{eqnarray}\label{sol2DKK2}
u=\frac{3}{(\det{A})^2}
\Bigg(\frac{\det{A}}{2v_1^2}\Big(-2v_1|\alpha_1|^2_x
(1-\frac{1}{2v_1}\partial_{x}^{-1}|\alpha_1|^2)-
\alpha_1\alpha_{1x}\partial_x^{-1}\overline{\alpha_1}^2-\overline{\alpha}_1\overline{\alpha}_{1x}\partial_x^{-1}\alpha_1^2\Big)
-\nonumber\\
-\frac{1}{16 v_1^4}\Big(\alpha_1^2\partial_x^{-1}\overline{\alpha_1}^2+\overline{\alpha}_1^2\partial_x^{-1}\alpha_1^2+4v_1|\alpha_1|^2
(1-\frac{1}{2v_1}\partial_{x}^{-1}|\alpha_1|^2)\Big)^2\Bigg).
\end{eqnarray}

In the case of  kernel $R_0$ (\ref{kernel_satify_CRed2&CR 1to2}) of delta-functional type, with $p_k(\mu,\overline{\mu})=A_k\delta(\mu-\mu_k)$, $(k=1,\ldots,N)$, which satisfies the conditions (\ref{CRed2}) and (\ref{RC1 CRed2 1to2_p}), due to definitions (\ref{alpha_FP,beta_FP}), functional parameters $\alpha_{k}$ have the following form:
\begin{equation}\label{CRed2 1to2 N1 DeltaFunction}
\alpha_{k}=-2i A_k e^{F(\mu_{k})},\quad (k=1,\ldots,N).
\end{equation}
Such kernel leads to corresponding exact multiline soliton solutions.
In the simplest case of $N=1$ one obtains from (\ref{Determinant CRed2 1to2 N1}), (\ref{sol2DKK2}) via (\ref{CRed2 1to2 N1 DeltaFunction}) the exact one line soliton solution of the 2DKK equation,
\begin{eqnarray}\label{Solution CRed2 1to2 N1 DeltaFunction}
\det{A}=1+\frac{2a}{\mu_{1I}}e^{\varphi(x, y, t)}+\frac{a^2}{|\mu_{1}|^2}
\frac{\mu_{1R}^2}{\mu_{1I}^2}e^{2\varphi(x, y, t)}=\big(1+\frac{a}{\mu_{1I}}e^{\varphi(x, y, t)}\big)^2-
\frac{a^2}{|\mu_{1}|^2}e^{2\varphi(x, y, t)},\nonumber\\
u(x, y, t) = \frac{24 a \mu_{1I}e^{\varphi}}{(\det{A})^2}
\Bigg[\det{A}-\frac{2a \mu_{1I}e^{\varphi}}{|\mu_{1}|^2}\Bigg],
\end{eqnarray}
where $a=v^{-1}_1 |A_1|^2=\bar a$ is some real parameter, $\varphi(x, y, t):=F(\mu_{1})-F(\overline{\mu_{1}})$; for $\frac{a}{\mu_{1I}}=\frac{ |A_1|^2}{v_1\mu_{1I}}>0$, due to expression for $\det A$ in (\ref{Solution CRed2 1to2 N1 DeltaFunction}), this solution  is nonsingular.

\section{Solutions of  2DSK equation.}

In present section the classes of solutions with functional parameters are calculated for  2DSK equation (\ref{2DSK}). This equation has  condition of reduction (\ref{reduction 2DSK}) which is different from that one (\ref{reduction 2DKK}) for 2DKK equation. It is convenient to transform this condition of reduction to determinant form, more appropriate for calculations with 2DSK equation (\ref{2DSK}). Substituting coefficients $\chi_{-1}$ and $\chi_{-2}$ from (\ref{chi_-1,-2_FP}) into the condition of reduction (\ref{reduction 2DSK}) one obtains,
\begin{equation}\label{SK_RC_FP}
  \sum_{k,\,l=1}^{N}(\alpha_{lx}\beta_k)A_{kl}^{-1}-\Big(\sum_{k,\,l=1}^{N}\frac{\partial A_{lk}}{\partial x}A_{kl}^{-1}\Big)^2=0.
\end{equation}
Defining degenerate matrix $V$ with elements $V_{lk} = \alpha_{lx}\beta_k$ and rank equal to unity, $\rank V=1$,
one rewrites condition (\ref{SK_RC_FP}) in the form,
\begin{equation}\label{SK_RC_tr}
  Tr(V\,A^{-1})-\Big[Tr\Big(\frac{\partial A}{\partial x}A^{-1}\Big)\Big]^2 = 0.
\end{equation}
By the use of identities (\ref{useful_identities_FP}) one obtains from (\ref{SK_RC_tr}) condition of reduction (\ref{reduction 2DSK}) in determinant form,
\begin{equation}\label{SK_RC_det}
  T(S-T)-T_x^2 = 0,
\end{equation}
where $T = \det{A}$, $S = \det{(A + V)}$.

The conditions of reality (\ref{RC1})  and reduction (\ref{reduction 2DSK}) or (\ref{SK_RC_det}) impose some restrictions on functional parameters.
In order to satisfy these conditions the terms in  kernel $R_0$ (\ref{Kernel FP}) will be grouped by pairs,
\begin{equation}\label{Kernel FP1 pairs 2DSK}
R_{0}(\mu,\overline{\mu},\lambda,\overline{\lambda})=
\pi\sum\limits_{k=1}^{2N}f_k(\mu,\overline{\mu})g_k(\lambda,\overline{\lambda})=\pi\sum\limits_{k=1}^{N}\Big[p_k(\mu,\overline{\mu})
q_k(\lambda,\overline{\lambda})+\tilde{p}_k(\mu,\overline{\mu})\tilde{q}_k(\lambda,\overline{\lambda})\Big].
\end{equation}
It is convenient to define via (\ref{Kernel FP1 pairs 2DSK}) the following sets $f$ and $g$ of functions $f_n$ and  $g_n$, $n=1\ldots 2N$,
\begin{eqnarray}\label{CRed1_sets fg KK}
f:=(f_1,\ldots,f_{2N})=(p_1(\mu,\overline{\mu}),\ldots, p_N(\mu,\overline{\mu});\tilde{p}_1(\mu,\overline{\mu}),\ldots, \tilde{p}_N(\mu,\overline{\mu})),\nonumber\\
g:=(g_1,\ldots,g_{2N})=(q_1(\lambda,\overline{\lambda}),\ldots, q_N(\lambda,\overline{\lambda});\tilde{q}_1(\lambda,\overline{\lambda}),\ldots, \tilde{q}_N(\lambda,\overline{\lambda})).
\end{eqnarray}

One shows easily that for case of $N=1$, i.e. for kernel (\ref{Kernel FP1 pairs 2DSK}) of the $\overline{\partial}$-problem with one pair of terms, the condition of reduction (\ref{reduction 2DSK}) or (\ref{SK_RC_det}) is fulfilled under the choice,
$\alpha_{2}=i\,c_1^{-1}\partial_x^{-1}\beta_1 ,\; \beta_2 = ic_1\alpha_{1x}$, with some complex constant $c_1$.
In terms functional parameters in spectral representation, due to definitions (\ref{alpha_FP,beta_FP}), last relation is rewritten equivalently  for the choice,
$\tilde{p}_1(\mu,\overline{\mu})=c_1^{-1}\mu^{-1}q_1(-\mu,-\overline{\mu}),\;
\tilde{q}_1(\lambda,\overline{\lambda}) = c_1\lambda p_1(-\lambda,-\overline{\lambda})$. By the use of symbolic calculations it was verified that the condition of the reduction (\ref{reduction 2DSK}) or  (\ref{SK_RC_det}) is satisfied for two such pairs of terms (i.e. $N=2$) in the kernel $R_0$ (\ref{Kernel FP1 pairs 2DSK}).
Generalizing last observation to the case of $N>2$ pairs of terms in (\ref{Kernel FP1 pairs 2DSK}) one choose the following sets of functional parameters in coordinate and in spectral representations  relating to each other by expressions,
\begin{eqnarray}\label{SK_N=4_beta_from_alpha}
   &\alpha_{k+N}&=i\,c_{k}^{-1}\partial_x^{-1}\beta_{k},\quad \beta_{k+N} = ic_{k}\alpha_{kx},\quad k = 1,\ldots,\,N,
   \nonumber\\
   &\tilde{p}_k(\mu,\overline{\mu})&=c_{k}^{-1}\mu^{-1}q_k(-\mu,-\overline{\mu}),\quad
\tilde{q}_k(\lambda,\overline{\lambda}) = c_{k}\lambda p_k(-\lambda,-\overline{\lambda}),
\end{eqnarray}
where $c_k, (k=1,..,N)$ are some complex constants and index $k$ numerates the pair of terms in kernel $R_0$;  for such choice the condition of reduction (\ref{reduction 2DSK}) or (\ref{SK_RC_det})  is fulfilled for the choice of interrelations between functional parameters of each pair in (\ref{Kernel FP1 pairs 2DSK}) given  by expressions (\ref{SK_N=4_beta_from_alpha}).

So due to (\ref{SK_N=4_beta_from_alpha}) the condition of reduction (\ref{reduction 2DSK}) or  (\ref{SK_RC_det}) is satisfied by choosing   kernel $R_0$ (\ref{Kernel FP1 pairs 2DSK}) of $\overline{\partial}$-problem (\ref{dibar problem}) in the following form,
\begin{equation}\label{SK_N=1_R0_RC}
  R_0(\mu,\overline{\mu},\lambda,\overline{\lambda}) =\pi \sum\limits_{k=1}^{2N}f_k(\mu,\overline{\mu})g_k(\lambda,\overline{\lambda})
=\pi \sum\limits_{k=1}^{N}\big[p_k(\mu,\overline{\mu})q_k(\lambda,\overline{\lambda}) + \frac{\lambda}{\mu} q_k(-\mu,-\overline{\mu})p_k(-\lambda,-\overline{\lambda})\big].
\end{equation}
In accordance with (\ref{SK_N=1_R0_RC})  the  sets $f$ and $g$ of functions $f_n$, $g_n$, $n=1,\ldots,2N$ in (\ref{CRed1_sets fg KK}) are taken the forms,
\begin{eqnarray}\label{CRed1_sets fg SK}
f:=(f_1,\ldots,f_{2N})=(p_1(\mu,\overline{\mu}),\ldots,
p_N(\mu,\overline{\mu});\frac{1}{\mu} q_1(-\mu,-\overline{\mu}),\ldots, \frac{1}{\mu} q_N(-\mu,-\overline{\mu})),\nonumber\\
g:=(g_1,\ldots,g_{2N})=(q_1(\lambda,\overline{\lambda}),\ldots, q_N(\lambda,\overline{\lambda}); \lambda p_1(-\lambda,-\overline{\lambda}),\ldots, \lambda p_N(-\lambda,-\overline{\lambda})).
\end{eqnarray}

The first condition of reality from (\ref{RC1}) is satisfied by imposing on each pair of terms in  sum (\ref{SK_N=1_R0_RC}) the following restrictions,
\begin{eqnarray}\label{kernel RC1 SK}
p_k(\mu,\overline{\mu})q_k(\lambda,\overline{\lambda}) + \frac{\lambda}{\mu} q_k(-\mu,-\overline{\mu})p_k(-\lambda,-\overline{\lambda})=\nonumber\\
=\overline{p_k(-\overline{\mu},-\mu)}\;\overline{q_k(-\overline{\lambda},-\lambda)} + \frac{\lambda}{\mu} \overline{q_k(\overline{\mu},\mu)}\;\overline{p_k(\overline{\lambda},\lambda)}.
\end{eqnarray}
Due to (\ref{kernel RC1 SK}) two cases are possible,
\begin{equation}\label{RC1 1to1 SK A}
A.\quad p_k(\mu,\overline{\mu})q_k(\lambda,\overline{\lambda})=
\overline{p_k(-\overline{\mu},-\mu)}\;\overline{q_k(-\overline{\lambda},-\lambda)},
\end{equation}
\begin{equation}\label{RC1 1to1 SK B}
B.\quad p_k(\mu,\overline{\mu})q_k(\lambda,\overline{\lambda})=
 \frac{\lambda}{\mu} \overline{q_k(\overline{\mu},\mu)}\;\overline{p_k(\overline{\lambda},\lambda)}.
\end{equation}

In the case $A$ by separating variables,
\begin{equation}\label{RC1 1to1 separate SK}
\frac{p_k(\mu,\overline{\mu})}{\overline{p_k(-\overline{\mu},-\mu)}}=
\frac{\overline{q_k(-\overline{\lambda},-\lambda)}}{q_k(\lambda,\overline{\lambda})}=v_k,
\end{equation}
with some complex constants $v_k$, $(k=1,...,N)$, one obtain  following restrictions on the functions $p_k(\mu,\overline{\mu})$ and $q_k(\lambda,\overline{\lambda})$,
\begin{equation}\label{RC1 1to1 p SK}
p_k(\mu,\overline{\mu})=v_k\overline{p_k(-\overline{\mu},-\mu)},\quad
q_k(\lambda,\overline{\lambda})=v_k^{-1}\overline{q_k(-\overline{\lambda},-\lambda)},\quad |v_k|^2=1.
\end{equation}

In the case $B$ by separating variables,
\begin{equation}\label{RC1 1to1_separate SK}
\frac{\mu p_k(\mu,\overline{\mu})}{\overline{q_k(\overline{\mu},\mu)}}=
\frac{\lambda\overline{p_k(\overline{\lambda},\lambda)}}{q_k(\lambda,\overline{\lambda})}=v_k,
\end{equation}
with some real constants $v_k$, $(k=1,...,N)$, one obtain  another restrictions on the functions $p_k(\mu,\overline{\mu})$,
\begin{equation}\label{RC1 1to1_p SK}
q_k(\mu,\overline{\mu})=\frac{\mu}{v_k} \overline{p_k(\overline{\mu},\mu)},\quad
v_k=\overline{v}_k=v_{k0}.
\end{equation}

Second condition of reality from (\ref{RC1}) for  2DSK equation (\ref{2DSK}) will be satisfied by
 imposing on each terms in  sum (\ref{SK_N=1_R0_RC}) the following restriction,
\begin{eqnarray}\label{kernel RC2 SK}
p_k(\mu,\overline{\mu})q_k(\lambda,\overline{\lambda}) + \frac{\lambda}{\mu} q_k(-\mu,-\overline{\mu})p_k(-\lambda,-\overline{\lambda})=\nonumber\\
=\overline{p_k(\overline{\lambda},\lambda)}\;\overline{q_k(\overline{\mu},\mu)} + \frac{\mu}{\lambda} \overline{q_k(-\overline{\lambda},-\lambda)}\;\overline{p_k(-\overline{\mu},-\mu)}.
\end{eqnarray}
Due to (\ref{kernel RC2 SK}), also as for (\ref{kernel RC1 SK}), two cases are possible,
\begin{eqnarray}\label{RC2 1to1 SK_1}
A'.\quad p_k(\mu,\overline{\mu})q_k(\lambda,\overline{\lambda})=
\overline{p_k(\overline{\lambda},\lambda)}\;\overline{q_k(\overline{\mu},\mu)},\nonumber\\
\frac{\lambda}{\mu} q_k(-\mu,-\overline{\mu})p_k(-\lambda,-\overline{\lambda})=\frac{\mu}{\lambda} \overline{q_k(-\overline{\lambda},-\lambda)}\;\overline{p_k(-\overline{\mu},-\mu)}.
\end{eqnarray}
\begin{eqnarray}\label{RC1 1to2 SK_1}
B'.\quad p_k(\mu,\overline{\mu})q_k(\lambda,\overline{\lambda})=
 \frac{\mu}{\lambda} \overline{q_k(-\overline{\lambda},-\lambda)}\;\overline{p_k(-\overline{\mu},-\mu)},\nonumber\\
 \frac{\lambda}{\mu} q_k(-\mu,-\overline{\mu})p_k(-\lambda,-\overline{\lambda})=
\overline{p_k(\overline{\lambda},\lambda)}\;\overline{q_k(\overline{\mu},\mu)}.
\end{eqnarray}

From  first case $A'$ (\ref{RC2 1to1 SK_1}) by separating variables one obtain the following expressions,
\begin{eqnarray}\label{RC2 1to1 SK}
\frac{p_k(\mu,\overline{\mu})}{\overline{q_k(\overline{\mu},\mu)}}=\frac{\overline{p_k(\overline{\lambda},\lambda)}}
{q_k(\lambda,\overline{\lambda})}=v_k, \nonumber\\
\frac{\lambda^2p_k(-\lambda,-\overline{\lambda})}{\overline{q_k(-\overline{\lambda},-\lambda)}}=
\frac{\mu^2\overline{p_k(-\overline{\mu},-\mu)}}{q_k(-\mu,-\overline{\mu})}=\tilde{v}_k,
\end{eqnarray}
with some complex constants $v_k$ and $\tilde{v}_k$
$(k=1,...,N)$. It follows from the last equation in
(\ref{RC2 1to1 SK}), due to the first equation from
(\ref{RC2 1to1 SK}), the relation
 $\lambda^2 v_k=\mu^2v_k=\tilde{v}_k$
which is impossible for arbitrary $\lambda$, $\mu$.

From  second case $B'$ (\ref{RC1 1to2 SK_1}) by separating variables one obtain the following expressions,
\begin{eqnarray}\label{RC2 1to2 SK}
\frac{p_k(\mu,\overline{\mu})}{\mu \overline{p_k(-\overline{\mu},-\mu)}}
=\frac{\overline{q_k(-\overline{\lambda},-\lambda)}}{\lambda q_k(\lambda,\overline{\lambda})}=v_k, \nonumber\\
\frac{\mu\overline{q_k(\overline{\mu},\mu)}}{q_k(-\mu,-\overline{\mu})}= \frac{\lambda p_k(-\lambda,-\overline{\lambda})}{\overline{p_k(\overline{\lambda},\lambda)}} =\tilde{v}_k,
\end{eqnarray}
with some complex constants $v_k$ and $\tilde{v}_k$ $(k=1,...,N)$.
It follows from the last equation in (\ref{RC2 1to2 SK}), due to the first equation from (\ref{RC2 1to2 SK}),
the relation, $\mu^2 v_k=-\tilde v_k$ which is impossible for arbitrary $\mu$.
So bellow  will be considered only the cases $A$ and $B$ defined by relations (\ref{RC1 1to1 SK A}) and (\ref{RC1 1to1 SK B}).

For  case $A$ (\ref{RC1 1to1 SK A}) the kernel $R_0$, which satisfies to conditions of reality (\ref{RC1}) and reduction (\ref{reduction 2DSK}) or (\ref{SK_RC_det}), has the form (\ref{SK_N=1_R0_RC}), where functions $p_k(\mu,\overline{\mu})$ and $q_k(\lambda,\overline{\lambda})$ are characterized by properties  (\ref{RC1 1to1 p SK}).
Due to definitions (\ref{alpha_FP,beta_FP}) and (\ref{SK_N=4_beta_from_alpha}), (\ref{RC1 1to1 p SK}) one derive the following
interrelations between different functional parameters,
\begin{equation}\label{alpha,beta CRed1 1to1 SK}
\alpha_k: = v_k\int\int\limits_C \overline{p_k(-\overline{\mu},-\mu)}e^{F(\mu)}d\mu \wedge d\overline{\mu}=-v_k\overline{\alpha_k},\quad
\beta_k: = \overline{v_k}\int\int\limits_C \overline{q_k(-\overline{\lambda},-\lambda)}e^{-F(\lambda)}
d\lambda \wedge d\overline{\lambda}=-\overline{v}_k\overline{\beta_k},
\end{equation}
\begin{equation}\label{alpha1,beta1 CRed1 1to1 SK}
\alpha_{k+N}: = c_k^{-1}\int\int\limits_C \frac{1}{\mu}q_{k}(-\mu,-\overline{\mu})e^{F(\mu)}d\mu\wedge d\overline{\mu}=i\,c_k^{-1}\partial_x^{-1}\beta_{k},\quad
\beta_{k+N}: = c_k\int\int\limits_C \lambda p_{k}(-\lambda,-\overline{\lambda})e^{-F(\lambda)}d\lambda \wedge d\overline{\lambda}=ic_k\alpha_{kx},
\end{equation}
where $|v_k|^2=1$ and $(k=1,\ldots,N)$.
So due to (\ref{alpha,beta CRed1 1to1 SK}), (\ref{alpha1,beta1 CRed1 1to1 SK}) the sets of functional parameters have the following structure,
\begin{equation}\label{set of alpha CRed2 1to1 SK}
(\alpha_1,\ldots,\alpha_{2N}):=(\alpha_1,\ldots,\alpha_N; i\,c_1^{-1}\partial_x^{-1}\beta_{1},\ldots,i\,c_N^{-1}\partial_x^{-1}\beta_{N}),
\end{equation}
\begin{equation}\label{set of beta CRed2 1to1 SK}
(\beta_1,\ldots,\beta_{2N}):=(\beta_1,\ldots,\beta_{N}; ic_1\alpha_{1x},\ldots, ic_N\alpha_{Nx}),
\end{equation}
i.e. both sets are expressed through $2N$ independent complex functional parameters $(\alpha_1,\ldots, \alpha_N)$ and $(\beta_1,\ldots, \beta_N)$ with properties (\ref{alpha,beta CRed1 1to1 SK}).

General determinant formula (\ref{Solution_general_formula_FP}) with matrix $A$ (\ref{matrix1_FP}) corresponding to the kernel $R_0$
(\ref{SK_N=1_R0_RC}) of the $\overline{\partial}$-problem (\ref{dibar problem}) due to (\ref{RC1 1to1 p SK}) gives the class of exact solutions $u$ with functional parameters of the 2DSK equation (\ref{2DSK}).
By construction due to (\ref{set of alpha CRed2 1to1 SK}), (\ref{set of beta CRed2 1to1 SK}) these solutions depend on $2N$ functional parameters $(\alpha_1,\ldots,\alpha_N)$ and $(\beta_1,\ldots,\beta_N)$.

In the simplest case $N=1$ $(\alpha_1,\alpha_{2}):=(\alpha_1,i\,c_1^{-1}\partial_x^{-1}\beta_{1})$,
$(\beta_1,\beta_2):=(\beta_1,ic_1\alpha_{1x})$  due to (\ref{alpha,beta CRed1 1to1 SK}), (\ref{alpha1,beta1 CRed1 1to1 SK}) the determinant of the matrix $A$  (\ref{matrix1_FP}) is given by expression,
\begin{eqnarray}\label{determinant CRed2 1to2 N1}
\det{A}=\Big(1-\frac{1}{4}\alpha_1\partial_{x}^{-1}\beta_1+\frac{1}{2}\partial_{x}^{-1}\alpha_1\beta_1\Big)^2=\Delta^2.
\end{eqnarray}
The corresponding solution $u$ is calculated with help of reconstruction formula (\ref{Solution_general_formula_FP}) and has the form,
\begin{equation}\label{Sol2DSK1}
    u = \frac{3}{2\Delta^2}\left[\Delta\left(\alpha_1\beta_{1x}-\alpha_{1xx}\partial_x^{-1}\beta_1\right)-\frac{1}{4}\left(\alpha_1\beta_1-
    \alpha_{1x}\partial_x^{-1}\beta_1\right)^2 \right]
\end{equation}
Due to  expression (\ref{determinant CRed2 1to2 N1}) for $\det A$ this solution is nonsingular for choices of functional parameters $\alpha_1,\beta_1$ satisfying to inequality, $-\frac{1}{4}\alpha_1\partial_{x}^{-1}\beta_1+\frac{1}{2}\partial_{x}^{-1}\alpha_1\beta_1>0$.

In the case of kernel $R_0$ (\ref{SK_N=1_R0_RC}) of delta-functional type with $p_k(\mu,\overline{\mu})=A_k\delta(\mu-i\mu_{k0})$, $q_k(\lambda,\overline{\lambda})=B_k\delta(\lambda-i\lambda_{k0})$, $k=1,\ldots,N$, which satisfy to conditions (\ref{SK_N=4_beta_from_alpha}) and
 (\ref{RC1 1to1 p SK}), due to definitions (\ref{alpha_FP,beta_FP}), the functional parameters have the following form,
\begin{equation}\label{CRed1 1to1 N1 DeltaFunction SK}
\alpha_{k}=-2i A_k e^{F(i\mu_{k0})},\quad
\beta_{k}=-2i B_k e^{-F(i\lambda_{k0})},
\end{equation}
where in accordance with (\ref{RC1 1to1 p SK}) $A_k=v_k \overline {A}_k$, $B_k=v_k \overline {B}_k$ and $\lambda_{k0}$, $\mu_{k0}$ - some real parameters. Such kernel leads to corresponding exact multiline soliton solutions.

In the simplest case of $N=1$  from (\ref{determinant CRed2 1to2 N1}), (\ref{Sol2DSK1}) due to (\ref{CRed1 1to1 N1 DeltaFunction SK}) one obtains, under the condition
$A_1 B_1\frac{\mu_{10}+\lambda_{10}}{\lambda_{10}(\mu_{10}-\lambda_{10})}=e^{\varphi_0}>0$, the exact nonsingular one line soliton solution of the 2DSK equation,
\begin{equation}\label{Solution CRed1 1to1 N1 DeltaFunction SK}
    u(x,y,t) = \frac{3 (\mu_{10}-\lambda_{10})^2}{2\cosh^2(\frac{\varphi+\varphi_0}{2})},
\end{equation}
where $\varphi = F(i\mu_{10}) - F(i\lambda_{10})$.
This one line soliton solution was derived earlier in the paper \cite{DubrLisit} of first author. For the case of $N=2$ one obtains the exact two line soliton solution of the 2DSK equation.

For  case $B$ (\ref{RC1 1to1 SK B}) the kernel $R_0$, which satisfies conditions of reality (\ref{RC1})  and reduction
(\ref{reduction 2DSK}) or (\ref{SK_RC_det}),  due to (\ref{SK_N=4_beta_from_alpha}), (\ref{SK_N=1_R0_RC}), (\ref{RC1 1to1_p SK}) has the form,
\begin{equation}\label{kernel 1B SK}
R_0(\mu,\overline{\mu},\lambda,\overline{\lambda}) =\pi \sum\limits_{k=1}^{2N}f_k(\mu,\overline{\mu})g_k(\lambda,\overline{\lambda})
=\pi \sum\limits_{k=1}^{N}\Big[v_k^{-1}\lambda p_k(\mu,\overline{\mu})\overline{p_k(\overline{\lambda},\lambda)} - v_k^{-1}\lambda \overline{p_k(-\overline{\mu},-\mu)}p_k(-\lambda,-\overline{\lambda})\big],
\end{equation}
where $v_k=v_{k0}$ - some real parameters.
From (\ref{kernel 1B SK}) one  choose convenient sets $f$ and $g$ of functions $f_n$, $g_n$, $n=1,..,2N$,
\begin{equation}\label{set of f CRed&CR 1to2 SK}
f:=(f_1,\ldots,f_{2N})=(p_1(\mu,\overline{\mu}),\ldots, p_N(\mu,\overline{\mu});-v_1^{-1} \overline{p_1(-\overline{\mu},-\mu)},\ldots,-v_N^{-1} \overline{p_N(-\overline{\mu},-\mu)}),
\end{equation}
\begin{equation}\label{set of g CRed&CR 1to2 SK}
g:=(g_1,\ldots,g_{2N})=(v_1^{-1}\lambda\overline{p_1(\overline{\lambda},\lambda,)},\ldots, v_N^{-1}\lambda\overline{p_1(\overline{\lambda},\lambda,)};
\lambda p_1(-\lambda,-\overline{\lambda}),\ldots, \lambda p_N(-\lambda,-\overline{\lambda})).
\end{equation}

Due to definitions (\ref{alpha_FP,beta_FP}) from (\ref{set of f CRed&CR 1to2 SK}), (\ref{set of g CRed&CR 1to2 SK}) one derive the following
interrelations between different functional parameters,
\begin{equation}\label{beta CRed 1to2 SK}
\beta_k: = v_k^{-1}\int\int\limits_C \lambda\overline{p_k(\overline{\lambda},\lambda,)}e^{-F(\lambda)}
d\lambda \wedge d\overline{\lambda}=-i v_k^{-1}\overline{\alpha}_{k\:x},\qquad k=1,\ldots,N
\end{equation}
\begin{equation}\label{alpha,beta1 CRed 1to2 SK}
\alpha_{k+N}: = -v_k^{-1}\int\int\limits_C \overline{p_k(-\overline{\mu},-\mu)}e^{F(\mu)}d\mu\wedge d\overline{\mu}=v_k^{-1}\overline{\alpha}_k,
\quad
\beta_{k+N}: = \int\int\limits_C \lambda p_k(-\lambda,-\overline{\lambda})e^{-F(\lambda)}d\lambda \wedge d\overline{\lambda}=i\alpha_{kx},
\end{equation}
where $v_k=v_{k0}$ - some real parameters.
So due to (\ref{beta CRed 1to2 SK}), (\ref{alpha,beta1 CRed 1to2 SK}) the sets of functional parameters have the following structure,
\begin{equation}\label{set of alpha CRed2 1to2 SK}
(\alpha_1,\ldots,\alpha_{2N}):=(\alpha_1,\ldots,\alpha_N; v_1^{-1}\overline{\alpha}_1,\ldots,v_N^{-1}\overline{\alpha}_N),
\end{equation}
\begin{equation}\label{set of beta CRed2 1to2 SK}
(\beta_1,\ldots,\beta_{2N}):=(-i v_1^{-1}\overline{\alpha}_{1x},\ldots,-i v_N^{-1}\overline{\alpha}_{Nx};
i\alpha_{1x},\ldots, i\alpha_{Nx}),
\end{equation}
i.e. both sets express through $N$ independent complex functional parameters $(\alpha_1,\ldots, \alpha_N)$.

General determinant formula (\ref{Solution_general_formula_FP}) with matrix $A$ (\ref{matrix1_FP}) corresponding to  kernel $R_0$
(\ref{kernel 1B SK}) of  $\overline{\partial}$-problem (\ref{dibar problem}) gives the class of exact solutions $u$ with functional parameters of  2DSK equation (\ref{2DSK}).
By construction, due to (\ref{set of alpha CRed2 1to2 SK}) and (\ref{set of beta CRed2 1to2 SK}), these solutions depend on $N$ complex functional parameters $(\alpha_1,\ldots,\alpha_N)$  given by (\ref{alpha_FP,beta_FP}).

In the simplest case $N=1$,
$(\alpha_1,\alpha_{2}):=(\alpha_1,v^{-1}_1\overline{\alpha}_1)$ and
$(\beta_1,\beta_2):=(-iv^{-1}_1\overline{\alpha}_{1 x},i\alpha_{1x})$,
the determinant of the matrix $A$ due to (\ref{matrix1_FP}) is given by expression,
\begin{eqnarray}\label{determinant CRed 1to2 N1 SK}
\det{A}=\big(1+\frac{i}{4v_1}\partial_{x}^{-1}(\alpha_{1x}\overline{\alpha}_1-
\alpha_1\overline{\alpha}_{1x})\big)^2=\Delta^2.
\end{eqnarray}
The corresponding solution $u$ is calculated with help of reconstruction formula (\ref{Solution_general_formula_FP}) and due to  (\ref{determinant CRed 1to2 N1 SK})  has the form:
\begin{equation}\label{sol2DSK2}
    u = \frac{3}{2 v_1 \Delta^2}\left[i\Delta\left(\alpha_{1xx}\overline{\alpha}_1-\alpha_1\overline{\alpha}_{1xx}\right)+\frac{1}{4 v_1}
    \left(\alpha_{1x}\overline{\alpha}_1-\alpha_1\overline{\alpha}_{1x}\right)^2\right].
\end{equation}
Due to  expression (\ref{determinant CRed 1to2 N1 SK}) for $\det A$ this solution is nonsingular for choices of functional parameter $\alpha_1$ satisfying to inequality, $\frac{i}{4v_1}\partial_{x}^{-1}(\alpha_{1x}\overline{\alpha}_1-
\alpha_1\overline{\alpha}_{1x})>0$.

In the case of kernel $R_0$ (\ref{SK_N=1_R0_RC}) of delta-functional type  with $p_k(\mu,\overline{\mu})=A_k\delta(\mu +\mu_{k})$, $k=1,\ldots,N$, which satisfy to conditions  (\ref{SK_N=4_beta_from_alpha}) and (\ref{RC1 1to1_p SK}), due to definitions from (\ref{alpha_FP,beta_FP}) functional parameters $\alpha_{k}$ have the following form,
\begin{equation}\label{CRed1 1to2 N1 DeltaFunction SK}
\alpha_{k}=-2i A_k e^{F(\mu_{k})}.
\end{equation}
The  kernels $R_0$ with such kind of functional parameters (\ref{CRed1 1to2 N1 DeltaFunction SK}) lead to corresponding exact multiline soliton solutions.

In the simplest case of $N=1$ from (\ref{determinant CRed 1to2 N1 SK}), (\ref{sol2DSK2}) and due to (\ref{CRed1 1to2 N1 DeltaFunction SK}), under condition $\frac{|A_1|^2}{v_1}\frac{\mu_{1R}}{\mu_{1I}}=e^{\varphi_0}>0$, one obtains the exact nonsingular one line soliton solution of the  2DSK equation,
\begin{equation}\label{Solution CRed1 1to2 N1 DeltaFunction SK}
    u(x,y,t) = \frac{6 \mu_{1I}^2}{\cosh^2(\frac{\varphi+\varphi_0}{2})},
\end{equation}
where $\varphi = F(\mu_{1}) - F(\overline{\mu_{1}})$.
This one line soliton solution was derived earlier in paper \cite{DubrLisit} of the first author. In the case of $N=2$ one obtains the exact two line soliton solution of the 2DSK equation.

\section{Periodic solutions of 2DKK and 2DSK equations}

In this section there will be calculated
 also the periodic solutions of 2DKK (\ref{2DKK}) and 2DSK (\ref{2DSK}) equations as particular cases from corresponding solutions with functional parameters obtained in sections III and IV.

{\bf{Periodic solutions of 2DKK equation.}}
At first using only reduction condition (\ref{reduction 2DKK}) or (\ref{reduction1 2DKK})
one calculates via general formulas  (\ref{Solution_general_formula_FP}),
(\ref{matrix1_FP}) (without using reality conditions), taking in to account  (\ref{CRed1 ab}) or (\ref{CRed1 fg})
(condition I of reduction (\ref{reduction 2DKK})), complex solution of 2DKK equation.
This solution for the simplest case $N=1$ in (\ref{Kernel FP1 1to1}) has the form,
\begin{equation}\label{solution reduction}
u=\frac{3}{c_1}\frac{\alpha_1 \alpha_{1x}\big(1+\frac{1}{2c_1}\partial_{x}^{-1}\alpha_1^2\big)-\frac{1}{4c_1}\alpha_1^4}
{\big(1+\frac{1}{2c_1}\partial_{x}^{-1}\alpha_1^2\big)^2}.
\end{equation}
For delta-functional kernel $R_0$, with $f_1:=A_1\delta(\mu-\mu_1)$
one obtains $\alpha_{1}=-2i A_1 e^{F(\mu_{1})}$, and from (\ref{solution reduction})
it  follows for $u$ at this stage complex  expression,
\begin{equation}\label{solution reduction delta}
u=\frac{-12A^2\mu_1 e^{\varphi}}
{\big(1+\frac{A^2}{\mu_1}e^{\varphi}\big)^2},
\end{equation}
where $2F(\mu_1)=\varphi$, $\frac{iA_1^2}{c_1}=A^2$. Using (\ref{solution reduction delta}) one formulates the reality condition ($u=\bar{u}$),
\begin{equation}\label{reality CRed1}
\frac{-12A^2\mu_1 e^{\varphi}}
{\big(1+\frac{A^2}{\mu_1}e^{\varphi}\big)^2}=
\frac{-12\overline{A}^2\overline{\mu}_1e^ {\overline{\varphi}}}
{\big(1+\frac{\overline{A}^2}{\overline{\mu}_1}e^{\overline{\varphi}}\big)^2}.
\end{equation}
Under assumptions $\mu_1=\overline{\mu_1}=\mu_{10}$ and $A^2=|\mu_{10}|e^{i\phi_a}$ (where $\phi_a$ - is arbitrary real constant) condition (\ref{reality CRed1}) fulfils, consequently  the phase $\varphi=-\overline{\varphi}=i\phi=2i(\mu_1 x+\mu_1^3 y+9\mu_1^5 t)$ is pure imaginary. Doing by this way and imposing on (\ref{solution reduction delta}) formulated restrictions one obtain   singular periodic solutions of 2DKK equation,
\begin{equation}\label{periodic solution1 CRed1}
\mu_{10}>0:\quad    u(x,y,t) = \frac{-3 \mu_{10}^2}{\cos^2(\frac{\phi+\phi_a}{2})};\qquad \mu_{10}<0: \quad u(x,y,t) = \frac{-3 \mu_{10}^2}{\sin^2(\frac{\phi+\phi_a}{2})}.
\end{equation}

Another periodic solutions of 2DKK equation  will be derived by use condition II of reduction (\ref{CRed2}).
One obtains directly from (\ref{Solution_general_formula_FP}), due to (\ref{matrix1_FP}), the non real solution $u\neq \overline u$, which satisfies only to condition (\ref{CRed2}) but doesn't satisfies to condition of reality. This solution for the simplest case one pair of terms ($N=1$) in the kernel (\ref{Kernel FP1 pairs CRed2}) is derived by reconstruction formula (\ref{Solution_general_formula_FP}) and due to (\ref{CRed2}), where determinant of matrix $A$ has the form,
\begin{equation}\label{solution reduction2}
\det A=\big(1+\frac{1}{2}\partial_{x}^{-1}\alpha_1\beta_1\big)^2-
\frac{1}{4}\partial_x^{-1}\alpha_1^{2}\:\partial_x^{-1}\beta_1^{2}.
\end{equation}
For the choice $f_1:=A_1\delta(\mu-\mu_1)$ and $g_1:=B_1\delta(\lambda-\lambda_1)$,  due to definitions (\ref{alpha_FP,beta_FP}) $\alpha_{1}=-2i A_1 e^{F(\mu_{1})}$ and  $\beta_{1}=-2i B_1 e^{-F(\lambda_{1})}$, for $u$ one obtains ($a:=iA_1B_1$) the expression,
\begin{eqnarray}\label{solution reduction2 delta 2DKK}
\det{A}=1+\frac{4a}{\mu_{1}-\lambda_{1}}e^{\varphi(x, y, t)}+\frac{a^2}{\mu_{1}\lambda_{1}}
\Bigg(\frac{\mu_{1}+\lambda_{1}}{\mu_{1}-
\lambda_{1}}\Bigg)^2e^{2\varphi(x, y, t)},\nonumber\\
u(x, y, t) =- \frac{12 a (\mu_{1}-\lambda_{1})e^{\varphi}}{\det{A}^2}\Bigg[\det{A}+
\frac{a(\mu_{1}-
\lambda_{1})e^{\varphi}}{\mu_{1}\lambda_{1}}\Bigg],
\end{eqnarray}
where $\varphi=F(\mu_1)-F(\lambda_1)$.

The reality condition ($u=\bar{u}$) with requirement of imaginary phase $\varphi=-\overline{\varphi}=i\phi$ leads to following conditions on the parameters:
\begin{equation}\label{reality CRed2}
\mu_1=\overline{\mu_1}=\mu_{10}, \quad \lambda_1=\overline{\lambda_1}=\lambda_{10}=c\mu_{10},\quad
a=\pm \sqrt{c}\mu_{10}\left(\frac{1-c}{1+c}\right),
\end{equation}
where $c$ is real parameter.
When one imposes on (\ref{solution reduction2 delta 2DKK}) these restrictions one obtain the singular periodic solutions,
for $c>0$,
\begin{equation}\label{periodic solution1 CRed1}
    u(x,y,t) = \mp12\sqrt{c}\mu^2_{10}\frac{(1-c)^2}{(1+c)}\frac{( 2\cos \phi \pm \frac{1+c}{\sqrt{c}})}{( 2\cos \phi \pm\frac{4\sqrt{c}}{1+c})^2};
\end{equation}
and for $c<0$,
\begin{equation}\label{periodic solution1 CRed1_a}
    u(x,y,t) = \pm 12\sqrt{|c|}\mu^2_{10}\frac{(1-|c|)^2}{(1+|c|)}\frac{( 2\sin \phi \mp\frac{1+|c|}{\sqrt{|c|}})}{( 2\sin \phi \mp\frac{4\sqrt{|c|}}{1+|c|})^2},
\end{equation}
where $\phi=\mu_{10}(1-c) x+\mu_{10}^3(1-c^3)y+9\mu_{10}^5(1-c^5) t$.

Another possibility to satisfy the reality condition ($u=\bar{u}$) for u given by (\ref{solution reduction2 delta 2DKK}) with requirement of imaginary phase $\varphi=-\overline{\varphi}=i\phi$ leads to another conditions on the parameters,
\begin{equation}\label{reality CRed2 2}
\lambda_1=-\overline{\mu_1},\quad
a^2=\pm|\mu_1|^2\left(\frac{\mu_{1R}}{\mu_{1I}}\right)^2.
\end{equation}
When one imposes  these restrictions on (\ref{solution reduction2 delta 2DKK}) one obtains the nonsingular (for arbitrary complex constant $\mu_1$) periodic solutions; for the choice  $a=\pm|\mu_1|\left(\frac{\mu_{1R}}{\mu_{1I}}\right)$,
\begin{equation}\label{periodic solution2 CRed1}
    u(x,y,t) = \mp12|\mu_1|\frac{\mu_{1R}^2}{\mu_{1I}}\frac{\cos \phi \pm\frac{\mu_{1I}}{|\mu_1|}}{(\cos \phi \pm\frac{|\mu_1|}{\mu_{1I}})^2},
\end{equation}
and for the choice  $a=\pm i|\mu_1|\left(\frac{\mu_{1R}}{\mu_{1I}}\right)$,
\begin{equation}\label{periodic solution2 CRed1 a}
    u(x,y,t) = \pm12|\mu_1|\frac{\mu_{1R}^2}{\mu_{1I}}\frac{\sin \phi \mp\frac{\mu_{1I}}{|\mu_1|}}{(\sin \phi \mp\frac{|\mu_1|}{\mu_{1I}})^2}.
\end{equation}
where $\phi=(\mu_1+\overline{\mu_1}) x+(\mu^3_1+\overline{\mu^3_1}) y+(\mu^5_1+\overline{\mu^5_1}) t$.

{\bf{Periodic solutions of 2DSK equation.}}
At this point is shown how one calculates periodic solution $u$ for 2DSK equation.
The solution $u$ which satisfies only condition of reduction (\ref{reduction 2DSK}) or  (\ref{SK_RC_det})  for case of $N=1$ in (\ref{SK_N=1_R0_RC}) has the form (\ref{Solution_general_formula_FP}) with $\det A$ due to (\ref{matrix1_FP}) and (\ref{SK_N=4_beta_from_alpha}) given by expression,
\begin{equation}\label{solution reduction2 SK}
\det A=\Big(1-\frac{1}{4}\alpha_1\partial_{x}^{-1}\beta_1+\frac{1}{2}\partial_{x}^{-1}\alpha_1\beta_1\Big)^2=\Delta^2.
\end{equation}
For delta-functional $f_1:=A_1\delta(\mu-\mu_1)$ and $g_1:=B_1\delta(\lambda-\lambda_1)$  due to definitions (\ref{alpha_FP,beta_FP})  follow expressions for functional parameters, $\alpha_{1}=-2i A_1 e^{F(\mu_{1})}$ and  $\beta_{1}=-2i B_1 e^{-F(\lambda_{1})}$, and for $u$ one obtains ($a:=iA_1B_1$),
\begin{eqnarray}\label{solution reduction2 delta}
\det{A}=\Big[1+a\frac{\mu_{1}+\lambda_{1}}{\lambda_1(\mu_{1}-
\lambda_{1})}e^{\varphi(x, y, t)}\Big]^2,\nonumber\\
u(x, y, t) = \frac{-6 a (\mu^{2}_{1}-\lambda^{2}_{1})}
{\lambda_{1}\det A}e^{\varphi(x, y, t)},
\end{eqnarray}
where $\varphi=F(\mu_1)-F(\lambda_1)$. The reality condition ($u=\bar{u}$) with requirement of imaginary phase $\varphi=-\overline{\varphi}=i\phi$ leads to following conditions on the parameters,
\begin{equation}\label{reality CRed2}
\mu_1=\overline{\mu_1}=\mu_{10}, \quad \lambda_1=\overline{\lambda_1}=\lambda_{10},\quad
a=|a|e^{i\phi_a}=\left|\frac{\lambda_{10}(\mu_{10}-
\lambda_{10})}{\mu_{10}+\lambda_{10}}\right|e^{i\phi_a}.
\end{equation}
When one imposes on (\ref{solution reduction2 delta}) these restrictions one obtain the singular periodic solutions of 2DSK equation (\ref{2DSK}),
\begin{equation}\label{periodic solution1 CRed1}
\frac{\lambda_{10}(\mu_{10}-\lambda_{10})}{\mu_{10}+
\lambda_{10}}>0: \quad
u(x,y,t) = \frac{-3 (\mu_{10}-\lambda_{10})^2}{2\cos^2(\frac{\phi+\phi_a}{2})};\qquad
\frac{\lambda_{10}(\mu_{10}-\lambda_{10})}{\mu_{10}+\lambda_{10}}<0: \quad
u(x,y,t) = \frac{-3 (\mu_{10}-\lambda_{10})^2}{2\sin^2(\frac{\phi+\phi_a}{2})},
\end{equation}
where $\phi=(\mu_{10}-\lambda_{10})x+(\mu^3_{10}-\lambda^3_{10})y+9(\mu^5_{10}-\lambda^5_{10})t$ and $\phi_a$ - is arbitrary real constant.

Another possibility to satisfy the reality condition ($u=\bar{u}$) with requirement of imaginary phase $\varphi=-\overline{\varphi}=i\phi$ leads to another conditions on the parameters,
\begin{equation}\label{reality CRed2 2}
\lambda_1=-\overline{\mu}_1,\quad
a=|a|e^{i\phi_a}=\left|\frac{\mu_{1R}}{\mu_{1I}}\right||\mu_{1}|e^{i\phi_a}.
\end{equation}
By imposing on (\ref{solution reduction2 delta}) these restrictions one obtain the another singular periodic solutions of 2DSK equation (\ref{2DSK}),
\begin{equation}\label{periodic solution2 CRed1}
\frac{\mu_{1R}}{\mu_{1I}}>0:\quad
u(x,y,t) = \frac{-6 \mu^{2}_{1R}}{\cos^2(\frac{\phi+\phi_a}{2})};\qquad
\frac{\mu_{1R}}{\mu_{1I}}<0:\quad
u(x,y,t) = \frac{-6 \mu^{2}_{1R}}{\sin^2(\frac{\phi+\phi_a}{2})},
\end{equation}
where $\phi=(\mu_1+\overline{\mu_1}) x+(\mu^3_1+\overline{\mu^3_1}) y+(\mu^5_1+\overline{\mu^5_1}) t$ and $\phi_a$ - is arbitrary real constant.

\section{Conclusions and acknowledgements}
The integrable 2DKK  (\ref{2DKK}) and 2DSK (\ref{2DSK}) equations differ only by coefficients (25/2 -- for 2DKK, and 5 -- for 2DSK) at nonlinear terms $u_x u_{xx}$. These equations arise as special reductions (\ref{V for 2DKK,2DSK}), $v=\frac{1}{2}u_x$ -- for 2DKK equation and $v=0$ -- for 2DSK equation, of more general integrable nonlinear system of equations for the corresponding fields in general position.
Such reductions lead to some nonlinear constraints (\ref{reduction 2DKK}), (\ref{reduction 2DSK}) on the wave functions of corresponding linear auxiliary problems.

In the present paper by the use of $\overline\partial$-dressing method of Zakharov and Manakov it is shown how  nonlinear constraints on wave functions are satisfied. By this way new classes of solutions with functional parameters of considered equations were constructed, as particular cases some periodic solutions also were obtained.

 It is interesting to note  that in  the paper \cite{DubrGramol} the gauge-invariant formulation of integrable 2DKK-2DSK system of equations was given. It was shown that 2DKK and 2DSK equations admit formulation in terms of corresponding gauge invariants and these equations are gauge nonequivalent to each other. Both of considered in the present  paper equations have different dispersionless partners, the investigation of these partners and calculation of theirs exact solutions is also interesting task and will be considered elsewhere.

 It was shown also that Nizhnik-Veselov-Novikov (NVN) and modifed Nizhnik-Veselov-Novikov (mNVN) equations also admit gauge-invariant formulation \cite{DubrGramol}, but in contrast to the case of 2DKK and 2DSK equations, NVN and mNVN equations belong  to gauge-equivalent classes of integrable nonlinear equations, their solutions can be connected by Miura-type transformations.

This research work is supported: 1. by scientific Grant of fundamental researches of Novosibirsk State Technical University  (2010); 2. by the Grant (registration number 2.1.1/1958) of Ministry of Science and Education of Russia Federation via analytical departmental special programm "Development of potential of High School" (2009-2011); 3. international research Grant RFFR \verb|#| 09-01-92442-Kea (2009-2010).

\end{document}